\documentclass[fleqn,usenatbib]{mnras}

\usepackage{txfonts}

\usepackage[T1]{fontenc}
\usepackage{ae,aecompl}

\usepackage[normalem]{ulem}

\usepackage{graphicx}
\usepackage{tabularx}
\usepackage{subfig}
\usepackage{comment}

\usepackage{amssymb}	
\usepackage{epsfig}
\usepackage{upgreek}
\usepackage{mathptmx}
\usepackage{txfonts}
\usepackage[utf8]{inputenc}
\usepackage{dcolumn}
\usepackage{natbib}
\usepackage{enumerate}
\usepackage{longtable,lscape}
\usepackage{color}
\usepackage{grffile}
\usepackage{array,float}
\usepackage{multirow}
\usepackage{lscape}
\usepackage{hyperref}
\usepackage{pdflscape}
\usepackage{afterpage}
\usepackage{capt-of}
\usepackage{xargs}
\usepackage[dvipsnames]{xcolor}
\usepackage{adjustbox}
\usepackage{booktabs}

\newcolumntype{R}[2]{%
    >{\adjustbox{angle=#1,lap=\width-(#2)}\bgroup}%
    l%
    <{\egroup}%
}

\def\THCO {\hbox{$^{13}{\rm CO}$}}   

\def\THCO {$\mathrm{^{13}CO}$} 
\def\CEIO {$\mathrm{C^{18}O}$} 
\def\FORM {$\mathrm{H_2CO}$} 
\def\METH {$\mathrm{CH_3OH}$} 
\def\MECN {$\mathrm{CH_3CN}$} 

\newcolumntype{d}[1]{D{.}{\cdot}{#1}}

\newcolumntype{.}{D{.}{.}{-1}}

\newcommand{\inv}{\ensuremath{^{-1}}}

\newcommand{\msun}{M$_\odot$}

\newcommand{\mclump}{\emph{M}$_{\rm{clump}}$}

\newcommand{\vlsr}{$V_{\rm{LSR}}$}

\newcommand{\mum}{$\mu$m}

\newcommand{\kms}{\ensuremath{\textrm{km\,s}^{-1}}}

\newcommand{\hi}{H{\sc i}}
\newcommand{\hii}{H{\sc ii}}
\newcommand{\uchii}{UC\,H{\sc ii}}

\title[DGF and SFE Across the Galactic Disk]{SEDIGISM-ATLASGAL: Dense Gas Fraction and Star Formation Efficiency Across the Galactic Disk\thanks{This publication is based on data acquired with the Atacama Pathfinder Experiment (APEX), projects 092.F-9315 and 193.C-0584. APEX is a collaboration between the Max-Planck-Institut f\"ur Radioastronomie, the European Southern Observatory, and the Onsala Space Observatory.}\thanks{The full version of Tables\,1, 2 and 3, and Fig.\,3 are only available in electronic form at the CDS via anonymous ftp to cdsarc.u-strasbg.fr (130.79.125.5) or via http://cdsweb.u-strasbg.fr/cgi-bin/qcat?J/MNRAS/.}}
\author[J.\,S.\,Urquhart et al.]{J.\,S.\,Urquhart$^{1}$\thanks{E-mail: j.s.urquhart@gmail.com}, C.\,Figura$^{2}$, J.\,R.\,Cross$^{1}$, M.\,R.\,A.\,Wells$^{1}$,
T.\,J.\,T.\,Moore$^{3}$,
D.\,J.\,Eden$^{3}$, \and 
S.\,E.\,Ragan$^{4}$,
A.\,R.\,Pettitt$^{5}$,  A.\,Duarte-Cabral$^{4}$,
D.\,Colombo$^{6}$,
F.\,Schuller$^{7,6}$,
T.\,Csengeri$^{8}$,\and
M.\,Mattern$^{9,6}$,
H.\,Beuther$^{10}$,
K.\,M.\,Menten$^{6}$,
F.\,Wyrowski$^{6}$,
L.\,D.\,Anderson$^{11,12,13}$, \and 
 P.\,J.\,Barnes$^{14}$,
M.\,T.\,Beltr\'an$^{15}$,
S.\,J.\,Billington$^{1}$,
L.\,Bronfman$^{16}$,
A.\,Giannetti$^{17}$,\and
J.\,Kainulainen$^{18}$,
J.\,Kauffmann$^{19}$ ,
M.-Y.\,Lee$^{20}$,
S.\,Leurini$^{21}$, 
S.-N.\,X.\,Medina$^{6}$,\and
F.\,M.\,Montenegro$^{22}$,
M.\,Riener$^{10}$,
 A.\,J.\,Rigby$^{4}$,
A.\,S\'anchez-Monge$^{23}$,
P.\,Schilke$^{23}$,\and
E.\,Schisano$^{24}$,
A.\,Traficante$^{24}$,
M.\,Wienen$^{25}$\\
Affiliations can be found after the references.}

\date{Accepted XXX. Received YYY; in original form ZZZ}

\pubyear{2020}

\begin{document}

\label{firstpage}
\pagerange{\pageref{firstpage}--\pageref{lastpage}}

\maketitle

\begin{abstract}

{By combining two surveys covering a large fraction of the molecular material in the Galactic disk we investigate the role the spiral arms play in the star formation process. We have matched clumps identified by ATLASGAL with their parental GMCs as identified by SEDIGISM, and use these giant molecular cloud (GMC) masses, the bolometric luminosities, and integrated clump masses obtained in a concurrent paper to estimate the dense gas fractions (DGF$_{\rm gmc}=\sum M_{\rm clump}/M_{\rm gmc}$) and the instantaneous star forming efficiencies (i.e., SFE$_{\rm gmc} = \sum L_{\rm clump}/M_{\rm gmc}$). We find that the molecular material associated with ATLASGAL clumps is concentrated in the spiral arms ($\sim$60\,per\,cent found within $\pm$10\,\kms\ of an arm). We have searched for variations in the values of these physical parameters with respect to their proximity to the spiral arms, but find no evidence for any enhancement that might be attributable to the spiral arms.  The combined results from a number of similar studies based on different surveys indicate that, while spiral-arm location plays a role in cloud formation and \hi\ to H$_2$ conversion, the subsequent star formation processes appear to depend more on local environment effects.  This leads us to conclude that the enhanced star formation activity seen towards the spiral arms is the result of source crowding rather than the consequence of a any physical process.  }

\end{abstract}





\section{\label{intro}Introduction}

Although comparatively few in number, massive stars play a significant role in the development and evolution of their host  galaxy (\citealt{kennicutt2012}).  Their high luminosities and UV fluxes can drive strong stellar winds and lead to the production of \hii\ regions that can influence their local environments and new trigger star formation by compressing the surrounding molecular gas (collect and collapse or radiatively driven implosion, e.g., \citealt{whitworth1994, bertoldi1989}). Conversely, their feedback can limit star formation by dispersing much of their own natal gas and thus limit the total fraction of molecular gas that can be turned into stars. Massive stars can, therefore, play an important role in regulating star formation (e.g. \citealt{dale2008,dib2011, dib2013}). The heavy elements produced through nucleosynthesis are distributed to the interstellar medium (ISM) throughout their lives via stellar winds and by supernovae at the end of these stars' lives, enhancing the chemical content of the ISM, and allowing more complex molecules to form. 

For all their importance, the formation process of these stars is still poorly understood (see review by \citealt{motte2018}).  Their comparative rarity means that massive star-forming regions tend to be widely separated, placing them at greater distances from us than the numerous low-mass star-forming regions that can be studied in great detail. Furthermore, the short timescales associated with their collapse causes them to reach the main sequence while still enshrouded in their natal cocoons, impairing our ability to observe them until after most traces of their formation environment have long been dispersed. The time frames over which high-mass stars form is still an open question, with YSO and \uchii-region lifetimes of a few 10$^5$\,yrs (\citealt{mottram2011b,davies2011}) but with simulations indicating that material is drawn in from larger distances giving timescales of a million years (e.g., \citealt{padoan2019}; see also discussion on lifetimes by \citealt{motte2018}).  

Previous work has shown that there is little or no dependence of the mean star-formation properties and other physical parameters of molecular clouds and dense clumps  (e.g. surface densities, velocity dispersion  and level of Galactic shear) on their location in the main Galactic disk and, especially, on their proximity to spiral arms \citep{eden2012,eden2013,moore2012,dib2012,eden2015,ragan2016,ragan2018,rigby2019}. A detailed study by \citet{dib2012} of the region of the 1$^{\rm st}$ quadrant covered by the Galactic Ring Survey \citep{jackson2006} found no correlation between the DGF and SFE as a function of a Galactic cloud's proximity to spiral arms or the level of shear they experience. However, some dependencies on spiral arm locations are observed in the disk of nearby spiral galaxies, however, including a gradient in stellar age across the arms (e.g. \citealt{shabani2018}) that is consistent with the density wave theory. The results from density wave theory are not supported by the observations of M51, however, where there is no evidence for the onset of star formation merely in spiral arms (\citealt{schinnerer2017}). The dynamic associated to spiral arms may influence the star formation efficiency (SFE) in their vicinity such as in spurs (e.g., \citealt{meidt2013}). This difference may be a result of the difficulties to differentiate between spiral arm and arm objects in the Milky way, which is not an issue for observations of nearby face-on spiral galaxy studies.

There is a large variation in the cloud-to-cloud star-formation efficiency and dense-gas fraction, as measured by the ratios of integrated IR luminosity and dense-gas mass within clouds to total molecular-cloud masses (\citealt{moore2012, eden2012,csengeri2016_planck}).  The same studies also showed that cloud-to-cloud variations in these parameters predominate, with ratios ranging over two orders of magnitude that are consistent with being log-normal distributions (\citealt{eden2015}). This variation does not originate from the uncertainty associated with using infrared luminosity as a SF tracer, as a similar variation is observed when using a more direct SFR tracer such as YSO counts (\citealt{lada2010, kainulainen2014, zhang2019}). However, we note that there can be a systematic offset between SFRs determined from infrared measurements and star counts as revealed by the detailed study of NGC\,346 reported by \citet{hony2015} and so some of the variation observed could be dependent on the choice of tracer.

A constant SFE, when averaged over kpc scales, is {\bf probably} consistent with simple empirical star-formation-rate scaling relations such as as the Schmidt-Kennicutt ``law'' (\citealt{gao2004,lada2012}), but the dominance of cloud-to-cloud variations indicates that, if there are physical mechanisms that regulate SFE, they operate principally on the scale of individual clouds. In particular, spiral arms appear to play little part in regulating or triggering star formation once a molecular cloud has formed. The high concentrations of molecular gas found to be associated with spiral arms in our own Galaxy and in nearby spiral galaxies indicates that it is likely that the arms play a role in triggering the cloud formation (via spiral shock or their gravitational potential), and therefore play an indirect role in the star formation process by enhancing giant molecular cloud (GMC) formation by converting \hi\ to H$_2$ (\citealt{koda2016,wang2020}). 

In this paper we will use the results of two Galactic plane surveys to investigate variations in the DGF and SFE across the inner Galactic disk. We compare the properties of clumps identified in the APEX Telescope Large Area Survey of the Galaxy (ATLASGAL; \citealt{schuller2009_full}) with those of their host molecular clouds identified from the final calibrated data cubes resulting from the SEDIGISM survey (Structure, Excitation and Dynamics of the Inner Galactic Interstellar Medium; \citealt{schuller2017}, {\color{red}Schuller et al. accepted}). Analysis of the SEDIGISM  data towards all ATLASGAL clumps provide a strong consistency check on the velocities already assigned from the other surveys utilised in our previous work.

The structure of this paper is as follows: in Sect.\,\ref{sect:survey_descriptions} we provide a brief overview of the survey and the data products used in this work. The extracted profiles are fitted with Gaussian components to obtain measurements of the amplitude, velocity and line-width of molecular material associated with the ATLASGAL clumps. We describe this process in Sect.\,\ref{sect:profile_fitting}, as well as the criteria used to assign a velocity in cases where two or more molecular components are detected.  In Sect.\,\ref{sect:results}, we use the velocities obtained from the CO analysis and the giant molecular cloud (GMC) catalogue produced from the SEDIGISM cubes ({\color{red}Duarte-Cabral et al. accepted.}) to derive the GMC dense gas fraction and star formation efficiency, and use these to look for variations towards the spiral arms. In Sect.\,\ref{sect:discussion} we discuss our results and investigate the role the spiral arms play in the star-formation process. We summarise our main findings in Sect.\,\ref{sect:summary}.

\section{Survey descriptions}
\label{sect:survey_descriptions}

\subsection{ATLASGAL}

{\bf ATLASGAL  (\citealt{schuller2009_full, beuther2012})} is an unbiased 870-$\mu$m\ submillimetre survey covering 420\,sq.\,degrees of the inner Galactic plane. It was specifically designed to identify an unbiased sample of dense, high-mass clumps that includes examples of all embedded evolutionary stages in the formation of massive stars. This survey has identified $\sim$10\,000 clumps distributed across the inner Galactic plane (\citealt{contreras2013,urquhart2014_csc,csengeri2014}); these clumps have sizes of $\sim0.5$\,pc and masses $\sim$500\,\msun\ (\citealt{urquhart2018}). 

\begin{figure}
 
\includegraphics[width=0.49\textwidth, trim= 0cm 0cm 0cm 0cm, clip]{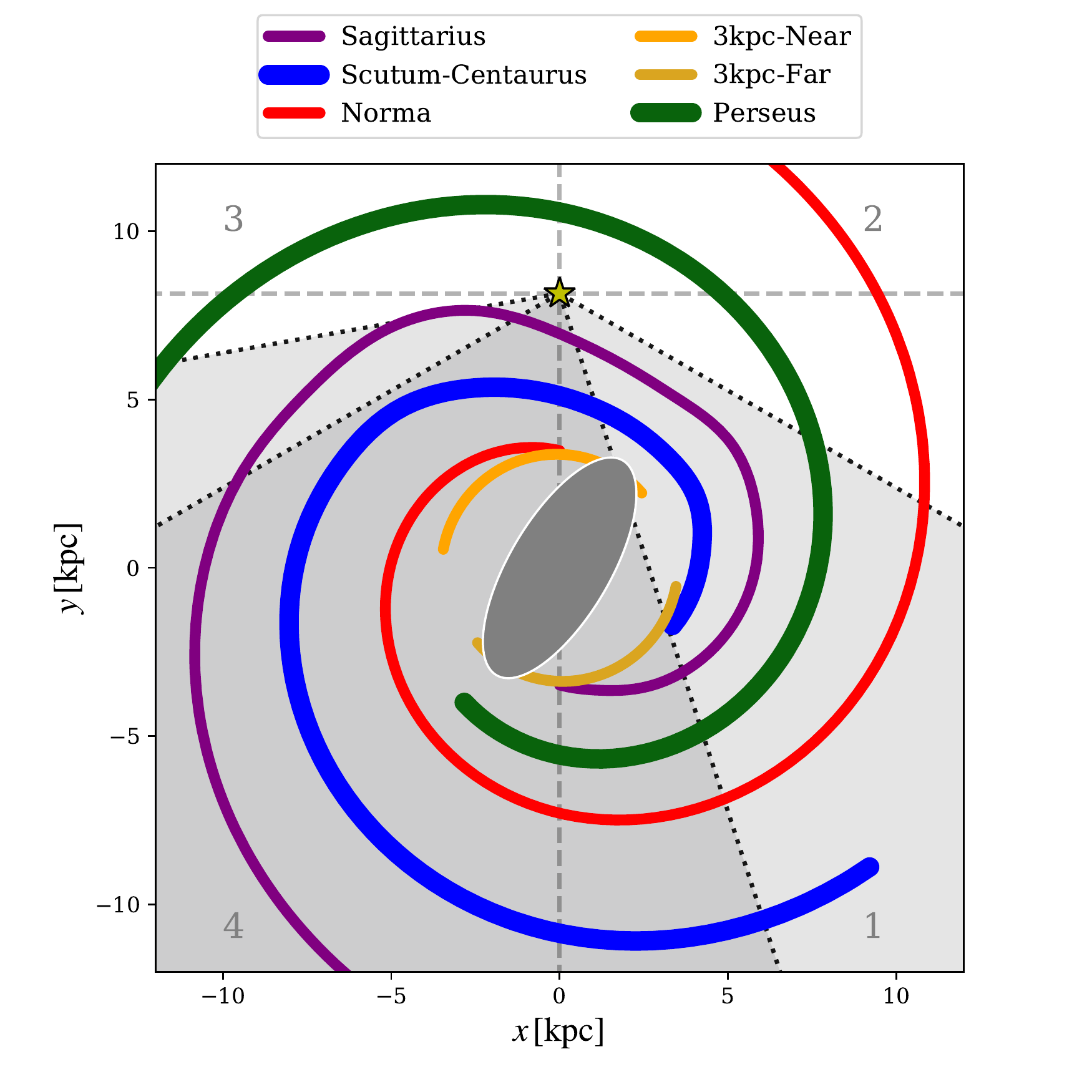}
  \caption{Schematic showing the loci of the spiral arms according to the model by \citet{taylor1993} and updated by \citet{cordes2004}, with an additional bisymmetric pair of arm segments added to represent the 3~kpc arms. The light grey-shaded area is the region covered by the ATLASGAL survey while the darker grey area indicates the region of the plane covered by the SEDIGISM survey. The star indicates the position of the Sun and the numbers identify the Galactic quadrants.  The bar feature is merely illustrative and does not play a role in our analysis. }
    \label{fig:topdown_view}
\end{figure}

 The area covered by this survey comprises $|\ell| < 60\degr$ with $|b|< 1.5\degr$ and $280\degr < \ell < 300\degr$ with $b$ between $-$2\degr\ and 1\degr. {\bf T}he shifted latitude was necessary to account for the warp in the Galactic disk in the outer Galaxy extension (see the light grey shaded region shown in Fig.\,\ref{fig:topdown_view}). In the current work we focus on the central part of the Galactic plane covered by both ATLASGAL and SEDIGISM (i.e. $300\degr< \ell < 18\degr$; see the dark shaded region shown in Fig.\,\ref{fig:topdown_view}).

 A crucial part of investigating the Galactic distribution and physical properties of dense, high-mass clumps is determining their distances. The most reliable distances are those determined from maser parallax measurements (e.g. \citealt{reid2019}); however, these are only readily available for approximately 200 regions (all with declinations $> -40\degr$), and although distance measurements are improving (particularly with respect to maser parallax measurements) it will be a long time before these will be available for a large fraction of the 10\,000 sources in the ATLASGAL  Compact Source Catalogue (CSC; \citealt{contreras2013,urquhart2014_csc}). Stellar parallax measurements from the Gaia mission \citep{gaia2018} are becoming more widely available (e.g., \citealt{zucker2019}), but these are primarily for regions of low extinction, and cannot be used to obtain distances to deeply-embedded protostars located at large distances in the dusty Galactic plane that are identified by ATLASGAL. We have therefore resorted to using kinematic distances for sources for which a more reliable distance measurement is not currently available. 

We presented velocities and kinematic distances for $\sim$8\,000 clumps located outside the Galactic centre region (i.e., $|\ell| > 5\degr$) in \citet{urquhart2018} based on the rotation curve of \citet{reid2014}. The radial velocities of molecular clumps can be measured from line observations (e.g., CO, NH$_3$, CS, etc.), and these are readily available for many of the ATLASGAL clumps.  We have used Galactic plane surveys such as the Galactic Ring Survey (GRS; \citealt{jackson2006}), the Mopra CO Survey of the Southern Galactic Plane (\citealt{burton2013,braiding2015}), ThrUMMS, (\citealt{barnes2015}), COHRS (\citealt{dempsey2013}), and CHIMPS (\citealt{rigby2016, rigby2019}) as well as large targeted observational programmes towards selected samples (e.g., MALT90 (\citealt{jackson2013}), RMS (\citealt{urquhart_13co_south,urquhart_13co_north, urquhart2011_nh3,urquhart2014_rms}), and BGPS (\citealt{dunham2011,schlingman2011_bgps_v, shirley2013}).  These have been augmented by dedicated ATLASGAL follow-up observations including NH$_3$ (\citealt{wienen2012, wienen2018}), SiO (2-1) \citep{csengeri2016_sio}, radio recombination lines \citep{kim2017, kim2018} and an unbiased 3-mm chemical survey between 85.2-93.4\,GHz \citep{urquhart2019}. Comparisons between the distances estimated by other survey teams and the ATLASGAL-determined distances finds agreement of $\sim$80\,per\,cent (see \citealt{urquhart2018} for more details).

The previous works had excluded the central part of the Galaxy because at the time there had been no high-resolution ($\le$1\arcmin) molecular line surveys or targeted studies that sufficiently resolved the complex source distributions. The recently-completed SEDIGISM survey (\citealt{schuller2017}) provides the means to extend the analysis of the ATLASGAL catalogue to the inner-most part of the Galactic plane and determine distances and physical properties for a significant number of the $\sim$2000 ATLASGAL sources currently without a distance. The vast majority of these are located towards the Galactic centre (i.e., $|\ell| < 5$\degr).

\begin{figure}
\centering 
\includegraphics[width=0.49\textwidth]{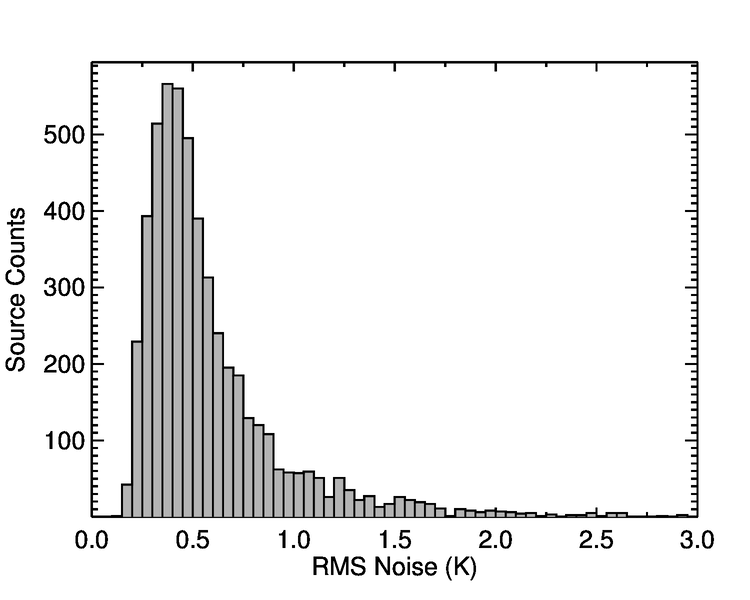}

\caption{Distribution of RMS noise values determined from the emission free regions of the C$^{18}$O (2-1) spectra smoothed to a velocity resolution of 1\,km\,s$^{-1}$. The bin size is 0.05\,K.}\label{fig:rms_dist}
\end{figure}

\subsection{SEDIGISM Survey}
\label{sect:sedigism_survey}

The SEDIGISM survey (\citealt{schuller2017} -- hereafter Paper\,I) utilised the 12-m Atacama Pathfinder Experiment  \citep[APEX,][]{gusten2006} submilllimeter telescope between 2013 and 2015 to observe the $J=2-1$ transitions of the \THCO\ and \CEIO\ isotopologues and 10 other significant molecular tracers, including shock tracers (such as SiO and SO) and dense gas tracers (\FORM, \METH, \MECN).

The observations used the Swedish Heterodyne Facility Instrument \citep[SHFI,][]{vassilev2008} paired with a back-end utilising two wide-band Fast Fourier Transform Spectrometers \citep[XFFTS,][]{Klein2012}.  Each spectrometer produced a 2.5\,GHz bandwidth with 32\,768 channels, yielding a velocity resolution of $\sim0.1\,\kms$ at the central frequency of the observations (219\,GHz).  The bands were configured to overlap by 500\,MHz, producing a net 4\,GHz IF bandwidth.

The survey area covers a 1-degree wide (in latitude) band over the southern Galactic plane ($-60\degr \le \ell \le +18$\degr, $|b| \le 0.5$\degr) with a 28-arcsec FWHM beam.  We divided this region into $0.5\times0.5$\,deg$^2$ fields, each of which was covered twice with orthogonal on-the-fly mapping. We used a 2-arcmin\,s\inv\ scanning speed to yield a $\sim0.34$\,s\,beam$^{-1}$ integration time.  When combined, these two mapping passes allow us to reach a main-beam brightness 1-$\sigma$ rms noise of 0.8\,K at 0.25\,\kms\ spectral resolution in typical weather conditions (maximum precipitable water vapour of 3\,mm).

A number of transitions are covered by the SEDIGISM survey, but the brightest are the $^{13}$CO and C$^{18}$O ($J =2-1$) rotational transitions. The $^{13}$CO  (2-1) transition requires a higher critical density than the $^{13}$CO (1-0) transition, and is less optically thick than the $^{12}$CO and $^{13}$CO (1-0) transitions. It is therefore a more reliable tracer of dense gas than the lower-excitation isotopologues, and is less affected by self-absorption and confusion due to blending of low-density clouds along the line of sight than lines from the much more abundant $^{12}$CO isotopologue. 

The $^{13}$CO and C$^{18}$O data are available in the form of 2\degr\ $\times$ 1\degr\ FITS cubes with a velocity range of $\pm200$\,\kms. These are centred on each integer value of Galactic longitude. These fits cubes are calibrated to the main beam temperature scale ($T_{\rm{mb}}$) and so have already been corrected for the APEX telescope beam efficiency ($\eta_{\rm eff} = 0.75$\footnote{http://www.apex-telescope.org/telescope/efficiency/index.php}). These fits cubes are available from the SEDIGISM project website.\footnote{http://sedigism.mpifr-bonn.mpg.de/} 

A detailed description of the whole survey and a discussion of the data quality and products are given in {\color{red} Schuller et al. accepted} (hearafter Paper\,II). This overview paper is complemented by a catalogue of GMCs ({\color{red}Duarte-Cabral et al. accepted.} -- hereafter Paper\,III) produced by applying the {\sc scimes} algorithm (v.0.3.2)\footnote{https://github.com/Astroua/SCIMES} (originally described in \citealt[][]{Colombo2015} with improvements detailed in \citealt[][]{Colombo2019}). This GMC catalogue consists of 10\,663 clouds and provides their physical parameters, such as distances, masses, sizes, velocity dispersions, virial parameters and surface densities{\bf . We  use} many of these parameters in Sect.\,\ref{sect:results}.

\begin{figure}
\centering 
\includegraphics[width=0.49\textwidth]{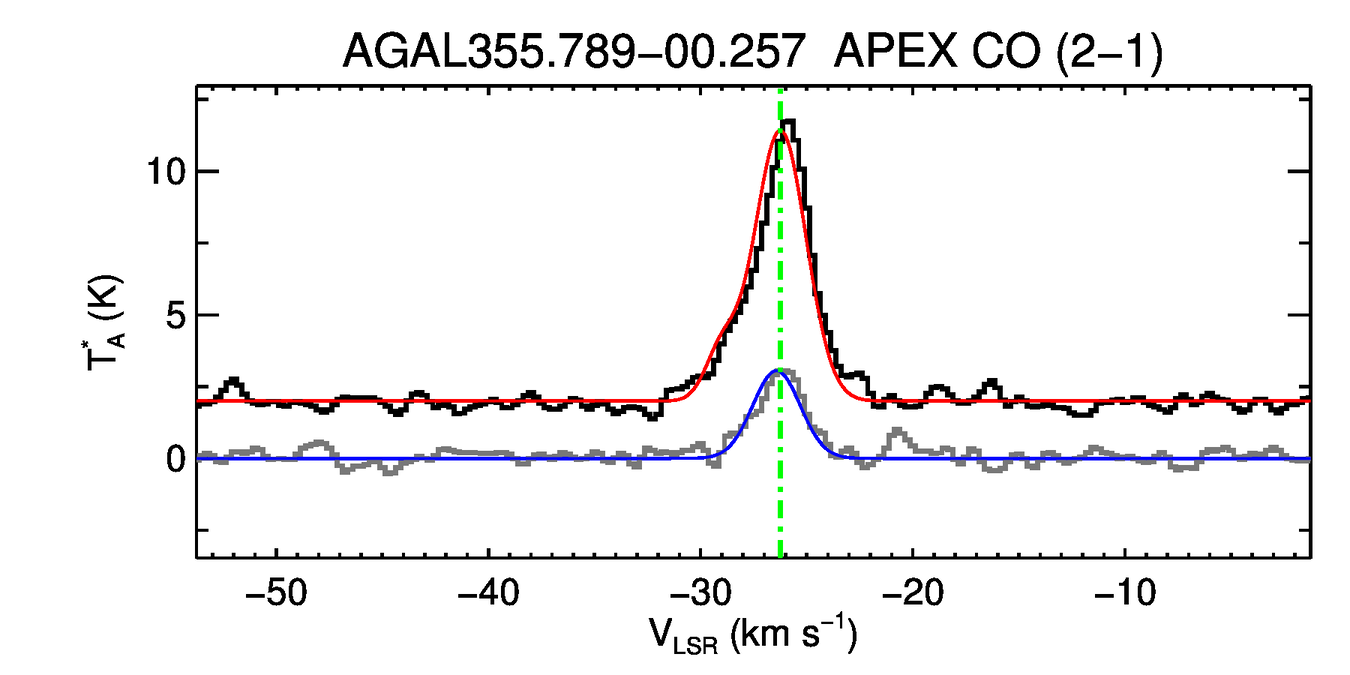}

\includegraphics[width=0.49\textwidth]{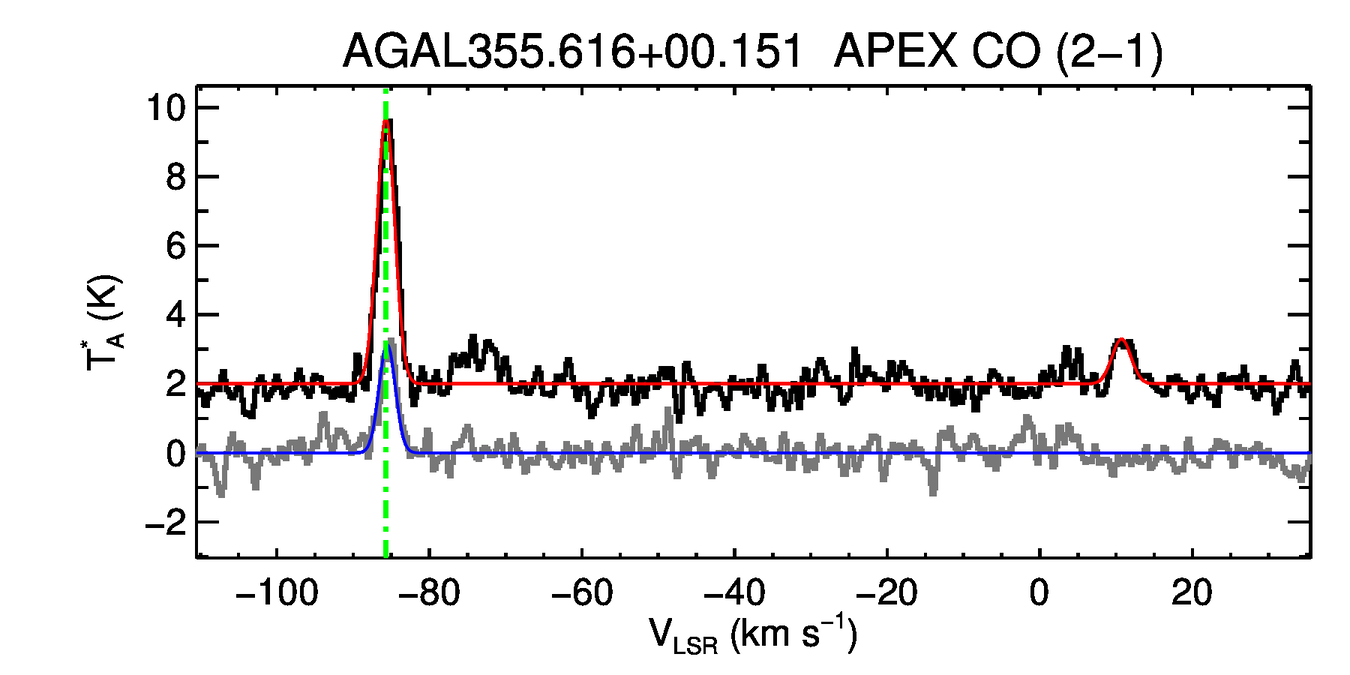}
\includegraphics[width=.49\textwidth, trim= 0 0 0 0]{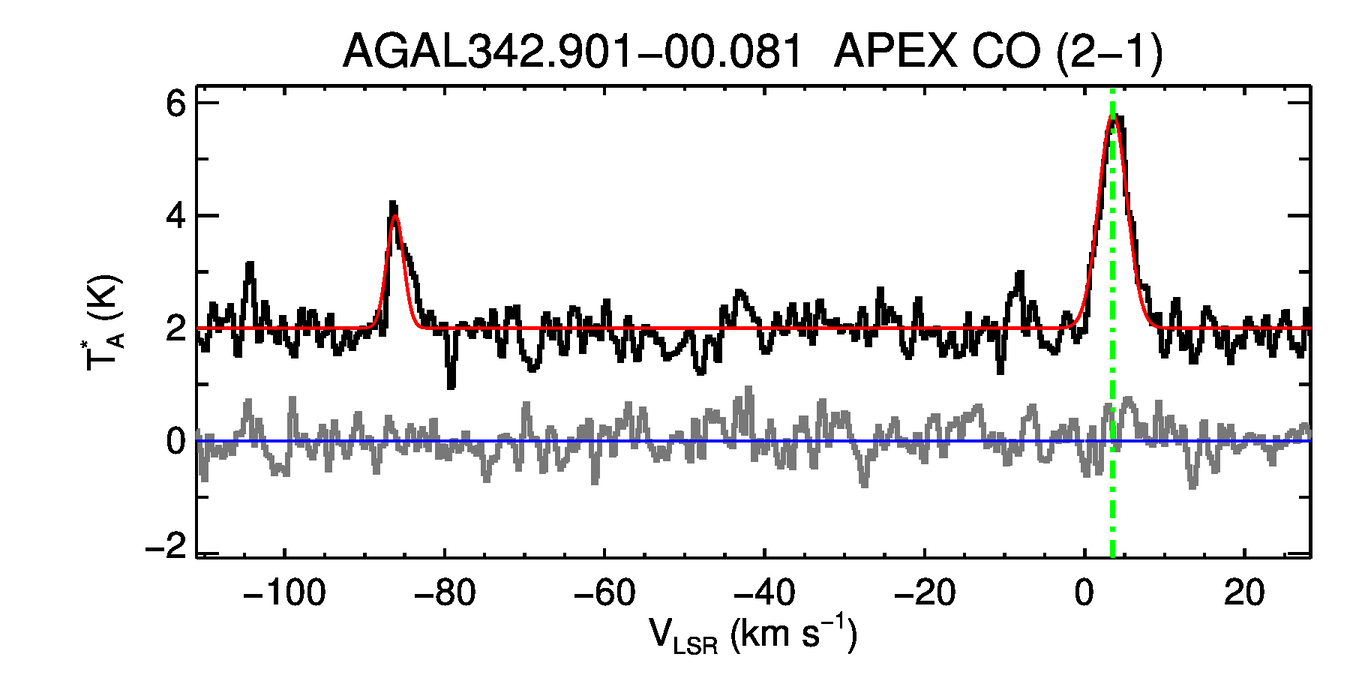}

\caption{Examples of the $^{13}$CO (black line) and  C$^{18}$O (grey) spectra extracted towards three dense clumps identified from the ATLASGAL survey; the former is offset by 2\,K from zero intensity. The results of the automatic Gaussian fitting are overlaid in red and blue and the velocity assigned to the clump is indicated by the green vertical dash-dotted line.}\label{fig:example_spectra}
\end{figure}

\section{$^{13}$CO and C$^{18}$O Analysis}
\label{sect:profile_fitting}

\subsection{Extraction and fitting}

There are 5754 clumps in the ATLASGAL CSC (\citealt{contreras2013,urquhart2014_csc}) that are located inside the region of the Galactic plane covered by SEDIGISM. We extracted spectra  for the $^{13}$CO and C$^{18}$O (2--1) transitions towards all of these positions by integrating the emission within a 30\arcsec\ aperture centred on the peak 870\,\mum\  dust emission. Inspection of the extracted $^{13}$CO data revealed that approximately 10\,per\,cent (606) of the spectra are affected by poor baselines and/or very broad emission features ($> 30$\,\kms; all of the latter are found towards the Galactic Centre).  We have excluded these from our analysis as they are unlikely to provide any reliable information for the clumps.

\begin{figure}
    \centering
    \includegraphics[width=.49\textwidth, trim= 0 0 0 0]{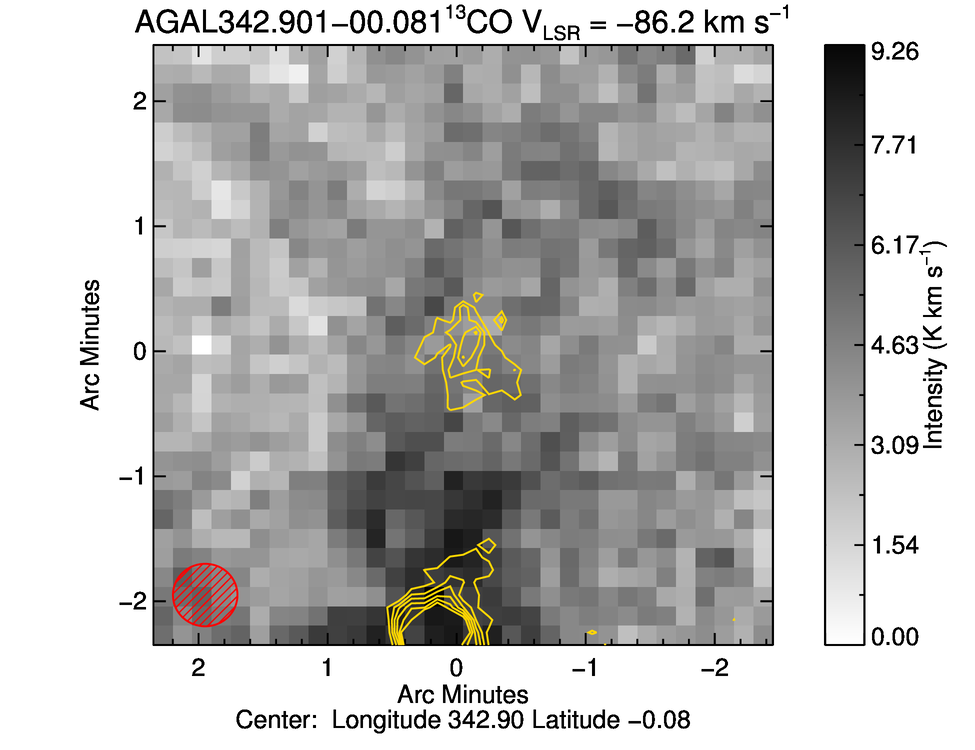}
    \includegraphics[width=0.49\textwidth, trim= 0 0 0 0]{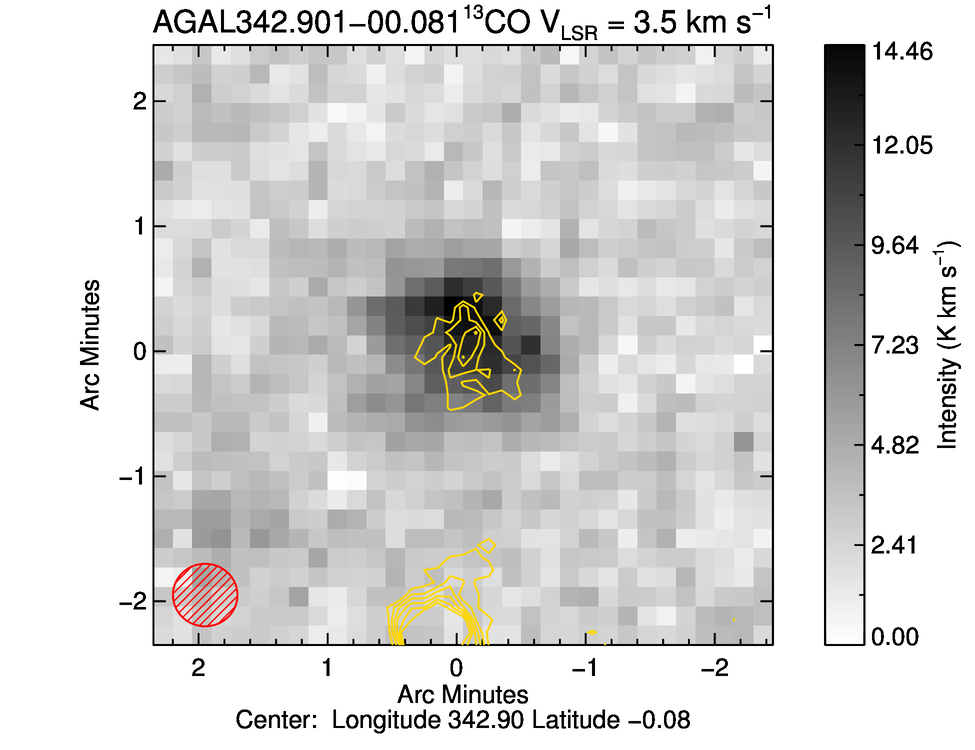}
\caption{Example of how integrated emission maps can be used to determine the most likely velocity component. The two maps above correspond to the two velocity components seen towards AGAL342.901$-$00.081 (see lower panel of Fig.\,\ref{fig:example_spectra}); the upper panel shows the integrated emission for the $\sim-86$\,\kms\ component while the lower panel shows the integrated emission at $\sim3.5$\,\kms. The contours show the distribution of the ATLASGAL 870\,\mum\ emission. The emission at $\sim$3.5\,\kms\ is strong, compact and correlated with the position of the ATLASGAL source and, therefore, is considered to be the most likely component. The SEDIGISM beam size is shown in the lower left corner of each map. }

\label{fig.integrated_maps}
\end{figure}

The spectral profiles towards the remaining 5148 clumps were Hanning smoothed, reducing the velocity resolution to 1\,\kms\ but improving the signal-to-noise ratio (SNR) by a factor of 2 ($\sigma=0.4$\,K; see Fig.\,\ref{fig:rms_dist} for distribution). The individual spectral components were automatically fitted assuming a Gaussian profile. The noise was estimated from emission-free regions of the spectrum. We then identified the strongest peak within the $\pm$200\,\kms\ velocity range: this was fitted, the Gaussian parameters were noted, and the fitted profile was subtracted from the spectrum. The next strongest peak was then identified, fitted and subtracted; this process was repeated until no peaks above 3$\sigma$ remained. A minimum threshold of 10\,per\,cent of the strongest component was employed to avoid over-complicating the analysis by taking data on very weak clouds that are very unlikely to be associated to the dense clumps identified in ATLASGAL. No attempt was made to separate the different components of strongly-blended emission or cases exhibiting self-absorption by simultaneously fitting multiple {\bf G}aussians. This simplification has the consequence of producing slightly larger uncertainties in a small number of cases.

\setlength{\tabcolsep}{2pt}
\begin{table}

\begin{center}\caption{
Fitted Gaussian parameters to CO spectra extracted towards ATLASGAL clumps.}
\label{tbl:co_fit_para}
\begin{minipage}{\linewidth}
\small
\begin{tabular}{lcccc}
\hline \hline
  \multicolumn{1}{l}{CSC Name}&  \multicolumn{1}{c}{Transition}&	\multicolumn{1}{c}{$T_{\rm mb}$}  &	\multicolumn{1}{c}{$v_{\rm lsr}$} &\multicolumn{1}{c}{FWHM} \\
  
    \multicolumn{1}{l}{}&  \multicolumn{1}{c}{}&	\multicolumn{1}{c}{(K)}  &	\multicolumn{1}{c}{(\kms)} &\multicolumn{1}{c}{(\kms)} \\
\hline
AGAL300.504$-$00.176	&	$^{13}$CO	&	10.75	$\pm$	0.11	&	8.51	$\pm$	0.03	&	4.48	$\pm$	0.10	\\
	&	$^{13}$CO	&	4.64	$\pm$	0.12	&	27.26	$\pm$	0.07	&	3.84	$\pm$	0.19	\\
	&	C$^{18}$O	&	1.64	$\pm$	0.12	&	8.51	$\pm$	0.20	&	4.02	$\pm$	0.58	\\
AGAL300.748$+$00.097	&	$^{13}$CO	&	20.74	$\pm$	0.14	&	$-$36.98	$\pm$	0.01	&	2.32	$\pm$	0.02	\\
	&	C$^{18}$O	&	5.32	$\pm$	0.16	&	$-$36.86	$\pm$	0.03	&	1.64	$\pm$	0.07	\\
AGAL301.136$-$00.226	&	$^{13}$CO	&	26.18	$\pm$	0.10	&	$-$39.62	$\pm$	0.02	&	5.92	$\pm$	0.06	\\
	&	C$^{18}$O	&	8.47	$\pm$	0.11	&	$-$39.61	$\pm$	0.04	&	4.67	$\pm$	0.14	\\
AGAL301.279$-$00.224	&	$^{13}$CO	&	7.28	$\pm$	0.13	&	$-$37.73	$\pm$	0.03	&	2.75	$\pm$	0.08	\\
	&	C$^{18}$O	&	2.78	$\pm$	0.17	&	$-$37.54	$\pm$	0.05	&	1.56	$\pm$	0.12	\\

\hline\\
\end{tabular}\\

{\bf Notes:} Only a small portion of the data is provided here, the full table will be available in electronic form at the CDS. 

\end{minipage}

\end{center}
\end{table}

\setlength{\tabcolsep}{6pt}

In total, 13\,117 $^{13}$CO and 5\,593 C$^{18}$O components are detected towards 5148 clumps.  The fitted line parameters are given in Table\,\ref{tbl:co_fit_para}. We present a few examples of the spectra obtained in Fig.\,\ref{fig:example_spectra}.

\subsection{Velocity determination}

\begin{figure}
    \centering
    \includegraphics[width=\columnwidth]{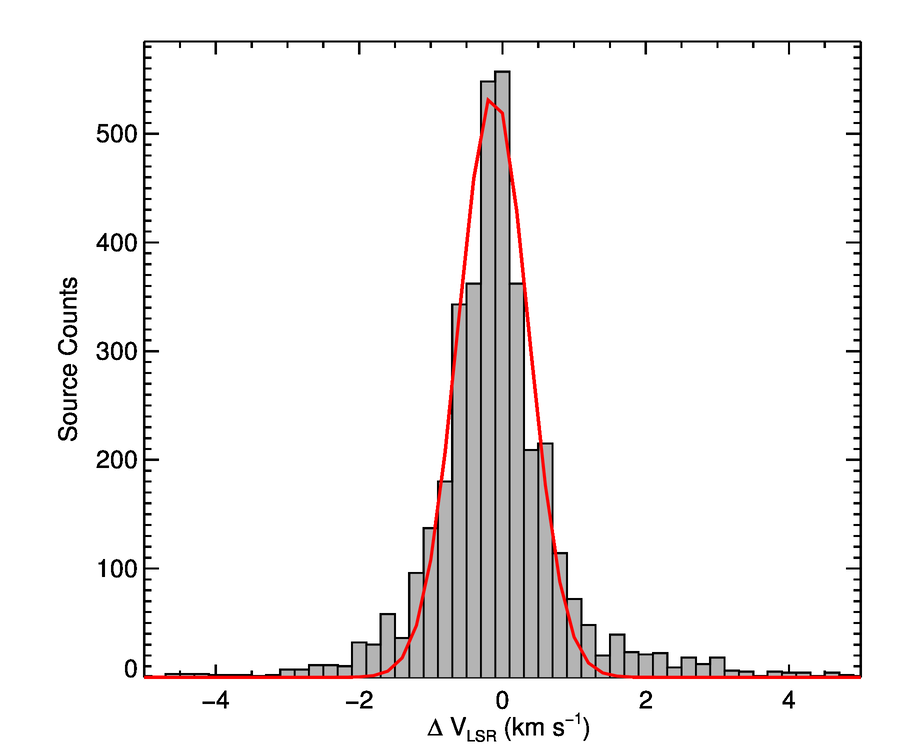}
    \caption{Histogram of the differences between previously assigned velocities (\citealt{urquhart2018}) and the velocities obtained from analysis of the SEDIGISM data discussed in this paper. The red curve shows the Gaussian fit to the distribution, which has a FWHM of 1.1\,\kms. The bin size is 0.2\,\kms.}
    \label{fig:SED_CSC_dV_histogram}
\end{figure}

Given that all of the ATLASGAL sources are located in the inner Galactic plane and that a significant fraction are located towards the Galactic centre, where the majority of molecular gas in the Galaxy resides, it is not surprising that multiple molecular components are found along the majority of sight-lines to the ATLASGAL clumps. A single component is detected in only $\sim$25\,per\,cent of cases (see top panel of Fig.\,\ref{fig:example_spectra}). Fortunately, in many of the multiple-detection cases there is one very strong peak that can be safely assumed to be the component associated with the dust emission (see middle panel of Fig.\,\ref{fig:example_spectra}). {\bf In} these cases the component with the largest integrated line intensity was allocated to the clump provided is at least twice the integrated intensity compared to the next-strongest component. Moreover, the detection of the C$^{18}$O line yields a clump's velocity unambiguously.

In other cases where the integrated intensities towards a particular clump are similar (i.e., within a factor of 2) we have produced integrated $^{13}$CO maps of $5\arcmin\times 5\arcmin$ regions centred on the peak dust emission (see lower panel of Fig.\,\ref{fig:example_spectra} for an example). These maps have been visually compared to the position and morphology of the dust emission, and the velocity of the integrated map that has the best morphological correlation is assigned to the clump. An example of this is shown in Fig.\,\ref{fig.integrated_maps} where the  integrated maps are presented for the two components seen towards the AGAL342.901$-$00.081 (see lower panel of Fig.\,\ref{fig:example_spectra}). It is clear from these maps that the velocity component at 3.5\,\kms\ is compact and is coincident with the position of the ATLASGAL clump, and we have assigned this velocity to the clump.

We have been able to assign a reliable velocity to 4998 clumps by selecting the strongest components and by examining the morphology of the CO emission with respect to the dust emission.  This results in 97\,per\,cent of the sample with useful extracted velocity data, of which 1108 velocities are newly-assigned.  The assigned velocities and CO-derived parameters are given in Table\,\ref{tbl:clump_derived_parameters} for all 4998 clumps with reliable velocities.

        \begin{figure*}
    \centering
    \includegraphics[width=\textwidth, trim= 10 0 10 0]{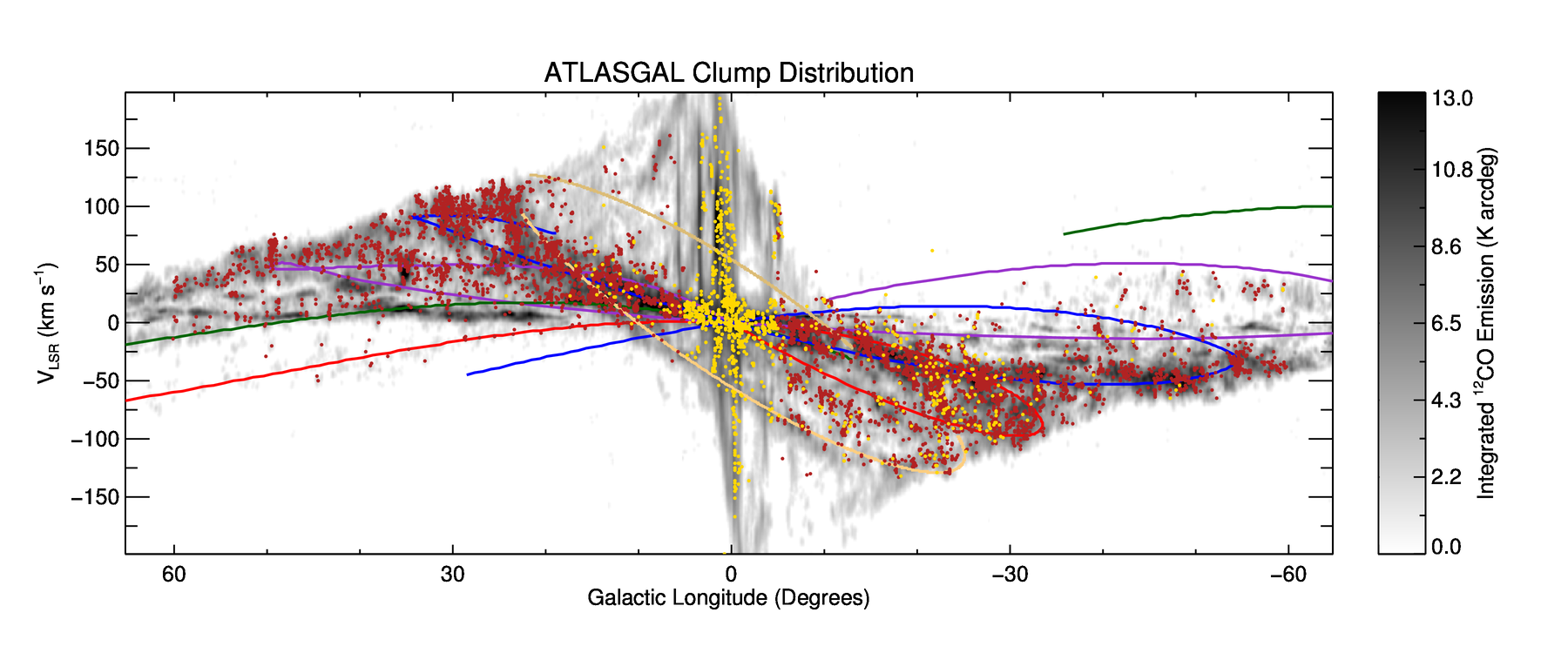}
\caption{Galactic longitude-velocity distribution of all ATLASGAL sources located between $300\degr < \ell < 60\degr$ for which we have been able to assign a velocity. The greyscale image shows the distribution of molecular gas as traced by the integrated $^{12}$CO (1-0) emission for comparison (\citealt{dame2001}). The red circles mark the positions of clumps where the velocity has been drawn from \citet{urquhart2018}, while the yellow circles mark the positions of clumps where a velocity has been determined from the work presented in this paper. The location of the spiral arms are shown as curved solid lines (colours are as given in Fig.\,\ref{fig:topdown_view}). The positions of the four main spiral and local arms have been taken from the model of \citet{taylor1993} and updated by \citet{cordes2004}.}

\label{fig.lv-distribution}
\end{figure*}

\setlength{\tabcolsep}{2pt}
\begin{table*}

\begin{center}\caption{Assigned velocities and derived physical parameters for the ATLASGAL clumps. The clumps' star formation efficiencies are taken directly from \citet{urquhart2018}.}
\label{tbl:clump_derived_parameters}
\begin{minipage}{\linewidth}
\small
\begin{tabular}{lc......c}
\hline \hline
  \multicolumn{1}{l}{CSC Name}&  \multicolumn{1}{c}{Transition}&	\multicolumn{1}{c}{$v_{\rm lsr}$}&	\multicolumn{1}{c}{$^{13}$CO($T_{\rm mb}$)}  & \multicolumn{1}{c}{C$^{18}$O($T_{\rm mb}$)}  &
  \multicolumn{1}{c}{FWHM$_{\rm ^{13}CO}$}  &\multicolumn{1}{c}{SFE$_{\rm clump}$} &\multicolumn{1}{c}{$R_{\rm gc}$} &	\multicolumn{1}{c}{GMC Name} \\
  
   \multicolumn{1}{l}{}&  \multicolumn{1}{c}{}&	\multicolumn{1}{c}{(\kms)}&	\multicolumn{1}{c}{(K)}  & \multicolumn{1}{c}{(K)}   &\multicolumn{1}{c}{(\kms)} & \multicolumn{1}{c}{($L_{\rm bol}$/ \mclump)} &
   \multicolumn{1}{c}{(kpc)}\\

\hline
AGAL341.126$-$00.347	&	$^{13}$CO	&	-41.4	&	14.6	&	8.0	&	4.32	&	8.68	&	5.07	&	SDG344.929$+$0.3022	\\
AGAL340.392$-$00.431	&	$^{13}$CO	&	-45.9	&	4.4	&	1.7	&	5.02	&	0.94	&	4.94	&	SDG343.133$-$0.4493	\\
AGAL340.311$-$00.436	&	$^{13}$CO	&	-48.1	&	5.5	&	3.1	&	4.31	&	0.52	&	4.85	&	SDG343.133$-$0.4493	\\
AGAL341.131$-$00.421	&	$^{13}$CO	&	-35.0	&	3.6	&	1.4	&	4.31	&	0.56	&	5.39	&	SDG344.257$-$0.3774	\\
AGAL340.349$-$00.434	&	$^{13}$CO	&	-47.9	&	3.9	&	1.5	&	6.31	&	0.48	&	4.86	&	SDG343.133$-$0.4493	\\
AGAL339.176$-$00.391	&	$^{13}$CO	&	-37.3	&	11.4	&	5.4	&	3.11	&	13.84	&	5.45	&	SDG341.016$-$0.1252	\\
AGAL339.403$-$00.414	&	$^{13}$CO	&	-39.1	&	4.5	&	1.9	&	3.37	&	1.19	&	5.34	&	SDG341.016$-$0.1252	\\
AGAL340.304$-$00.376	&	$^{13}$CO	&	-51.6	&	7.2	&	4.1	&	4.17	&	1.00	&	4.71	&	SDG343.133$-$0.4493	\\
AGAL340.269$-$00.416	&	$^{13}$CO	&	-49.1	&	5.6	&	2.3	&	3.72	&	0.65	&	4.81	&	SDG343.133$-$0.4493	\\
AGAL339.886$-$00.421	&	$^{13}$CO	&	-44.8	&	3.5	&	0.9	&	1.38	&	0.23	&	5.04	&	SDG342.364$+$0.0084	\\
\hline\\
\end{tabular}\\

{\bf Notes:} Only a small portion of the data is provided here: the full table will be available in electronic form at the CDS. 

\end{minipage}

\end{center}
\end{table*}

\setlength{\tabcolsep}{6pt}

\setlength{\tabcolsep}{3pt}
\begin{table}

\begin{center}\caption{Properties of matched SEDIGISM GMCs.}
\label{tbl:cloud_derived_parameters}
\begin{minipage}{\linewidth}
\small
\begin{tabular}{l....}
\hline \hline
  \multicolumn{1}{l}{GMC Name}&  \multicolumn{1}{c}{FWHM}&
  \multicolumn{1}{c}{$R_{\rm gc}$} & \multicolumn{1}{c}{DGF$_{\rm gmc}$}&	\multicolumn{1}{c}{SFE$_{\rm gmc}$} \\
  
   \multicolumn{1}{l}{}&	\multicolumn{1}{c}{(\kms)}&	
   \multicolumn{1}{c}{(kpc)} & \multicolumn{1}{c}{($\Sigma M_{\rm clump}/M_{\rm gmc}$)}   &\multicolumn{1}{c}{($\Sigma L_{\rm clump}$/$M_{\rm gmc}$)} \\
  
\hline
SDG348.053$+$0.2462	&	1.5	&		7.96	&	0.25	&	1.20	\\
SDG348.894$-$0.1875	&	1.8	&		9.56	&	0.14	&	7.83	\\
SDG349.776$+$0.0208	&	1.6	&		2.04	&	0.35	&	0.03	\\
SDG348.420$+$0.1106	&	0.7	&		2.24	&	0.64	&	0.13	\\
SDG349.240$+$0.0293	&	3.0	&		2.33	&	0.24	&	1.61	\\
SDG348.443$+$0.1729	&	1.1	&		2.53	&	0.33	&	0.25	\\
SDG349.108$+$0.0988	&	2.7	&		2.71	&	0.24	&	7.53	\\
SDG348.591$+$0.1546	&	1.0	&		2.69	&	0.67	&	0.57	\\
SDG349.805$+$0.0426	&	1.2	&		2.53	&	0.29	&	1.00	\\
SDG348.844$+$0.1346	&	2.3	&		2.92	&	0.33	&	0.37	\\
\hline\\
\end{tabular}\\

{\bf Notes:} Only a small portion of the data is provided here: the full table will be available in electronic form at the CDS. 

\end{minipage}

\end{center}
\end{table}

\setlength{\tabcolsep}{6pt}

In Figure\,\ref{fig:SED_CSC_dV_histogram} we present a histogram comparing the velocities obtained from our analysis of the SEDIGISM data with the previously assigned velocities (i.e., as described in \citealt{urquhart2018}). This plot shows the agreement between the two sets of independently assigned velocities, but also reveals that the velocities disagree by more than 3\,\kms\ for 269 sources, corresponding to $\sim$8\,per\,cent of the sample.\footnote{The choice of what value constitutes a significant difference is somewhat arbitrary.  Here, we have used a threshold difference value of three times the velocity resolution of the smoothed SEDIGISM data, which is 1\,\kms.} A more detailed investigation shows that the vast majority of these velocity disagreements were previously assigned using lower angular resolution $^{13}$CO\,(1-0) from the Mopra CO Survey of the Southern Galactic Plane (\citealt{burton2013,braiding2015}) or ThrUMMS (\citealt{barnes2015}). In these cases we consider the velocities assigned using the integrated $^{13}$CO\,(2-1) emission maps to be more reliable than using a single spectral profile, and so have adopted SEDIGISM velocities for these sources. We will use these new velocities to recalculate the distances for these 269 clumps and re-evaluate their cluster associations (these results will be presented in a subsequent paper).

Wherever possible, we adopt the velocities determined from SEDIGISM data for the rest of the clumps (3\,621), as this then provides a consistent set of molecular line fit parameters for a large fraction of the CSC catalogue and will allow for a more robust statistical analysis. The difference in the radial velocities is relatively modest (i.e., $<3$\,\kms) and so will not significantly affect the kinematic ambiguity distance solution, the kinematic distance or the clustering results presented in \citet{urquhart2018} and so we make no changes to any of the physical properties of these clumps presented in that paper.

\section{Star forming properties of host GMCs}
\label{sect:results}

In total, there are 5754 ATLASGAL clumps located in the SEDIGISM region and we have been able to allocate reliable velocities to 4998 clumps. We have used the positions and velocities of these clumps to match them to their parental GMCs as described in Paper\,III (see also Sect.\,\ref{sect:sedigism_survey}).  This has resulted in matching 4824 of the clumps with reliable velocities with  1709 GMCs, corresponding to 97\,per\,cent of dense clumps.

We note that only a small proportion of GMCs identified in the SEDIGISM data are associated with dense gas as traced by ATLASGAL ($\sim$11\,per\,cent, increasing to $\sim17$\,per\,cent in the disk).\footnote{We consider clouds located $|\ell| > 10$\degr\ to be located in the disk for this analysis.} Although the fraction of clouds with ATLASGAL counterparts is relatively small, they do make up approximately half of the total GMC mass.\footnote{In the $\ell = 300\degr -350\degr$ region there are 6352 GMCs with a total mass of 10$^{7.4}$\,\msun. Of these, 1044 are associated with an ATLASGAL clump and these GMCs have a combined mass of 10$^{7.1}$\,\msun.} The physical properties of the GMCs associated with dense clumps, and high-mass star formation tracers, were compared in Paper\,III together with the rest of the GMC population, and were found to be significantly more massive, physically larger in size, have higher velocity dispersion and surface densities; clouds associated with clumps had larger values and those associated with high-mass star forming tracers had even higher values (see figure\,8 of Paper\,III for distributions of the different populations).

In the rest of this section we look at the Galactic distribution of this sample of dense clumps and their host GMCs, and investigate whether their proximity to the spiral arms has any affect on their star-forming properties. For clarity, when we mention {\it clumps} we will always be referring to ATLASGAL sources, and when we mention {\it clouds} or {\it GMCs} we will always be referring to the SEDIGISM catalogue sources.

\subsection{Galactic distribution and association with spiral arms}

We show in Fig.\,\ref{fig.lv-distribution} the full distribution of ATLASGAL clumps that have an assigned velocity from the work presented both here and in \citet{urquhart2018}. This plot shows the longitudes and velocities for 8\,948 ATLASGAL CSC clumps of the  9\,817 located in the range $300\degr < \ell < 60\degr$, which corresponds to 91\,per\,cent of the catalogue. Two-thirds of the sources without an associated velocity are located within 3\degr\ of the Galactic centre, and their spectral profiles are too confused to extract a reliable velocity measurement.  Source confusion is compounded by the uncertainty of the rotation curve  toward the central region of the Galaxy, and so resulting kinematic distances would be highly uncertain.  The remaining third are among the weakest sources in the ATLASGAL CSC, and so there is a chance they are spurious. Our velocity information is therefore likely to be as complete as obtained. Figure\,\ref{fig.lv-distribution} also shows the loci of the four main spiral arms and the near/far 3-kpc arms.  The $x$ and $y$ positions of these arms have been taken from the model of \citet{taylor1993} as updated by \citet{cordes2004}, and have been converted to $\ell$ and \vlsr\ using a three-component rotation curve (bulge + disk + dark halo)  tailored to the data of \citet{eilers2019} and using the \citet{reid2019} values for Solar position and velocity (8.15\,kpc and 236\,\kms) and assuming pure circular rotation (as described in Paper\,II). The choice of rotation curve and spiral arm model do not make a significant difference to the spiral arm tracks on the $\ell-v$-map, as the differences in velocity are generally smaller than the streaming motions (this will be discussed in detail in a future publication i.e. Colombo et al. 2020, in prep.).

        \begin{figure}
    \centering
    \includegraphics[width=0.49\textwidth, trim= 0 0 0 0]{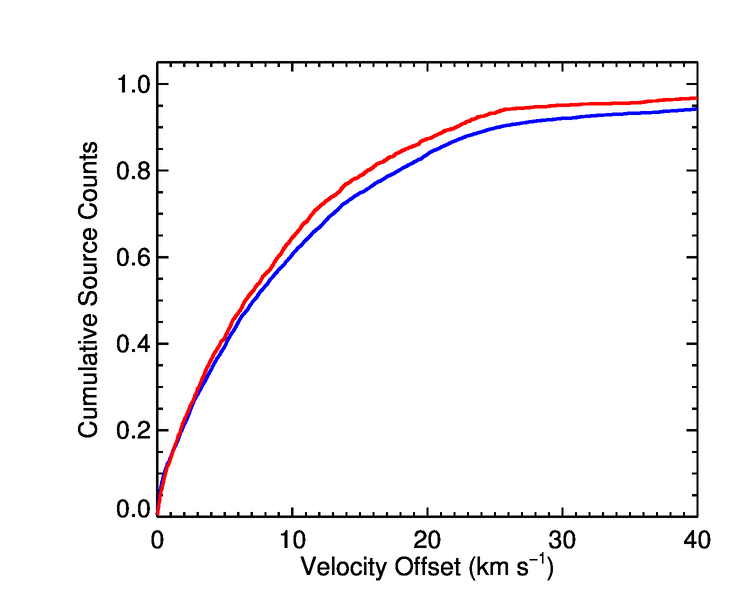}
     \includegraphics[width=0.49\textwidth, trim= 0 0 0 0]{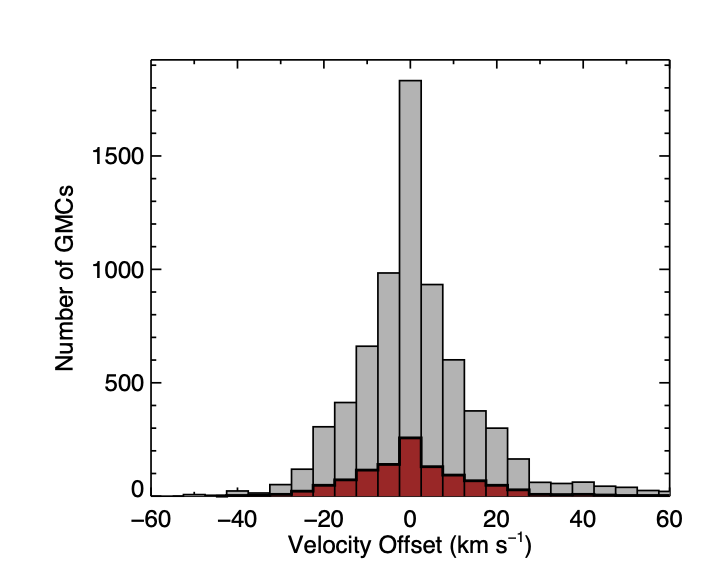}
\caption{The upper panel shows the cumulative distribution of the absolute velocity separations of all ATLASGAL clumps (red curve) and SEDIGISM GMCs (blue curve) from their nearest spiral-arm locus in $\ell-v$ space. The lower panel shows the frequency distribution of velocity differences of all GMCs (grey) and those associated with dense gas (red).}

\label{fig:arm_offset_vlsr}
\end{figure}

We can compare the distribution of the individual clumps with the loci of the spiral arms to look for trends in their physical and star-forming properties. In Fig.\,\ref{fig:arm_offset_vlsr} we show the velocity difference between the  clumps and {\bf the} nearest spiral arm in $\ell - v$-space for sources located beyond 10\degr\ of  longitude from the Galactic centre. The upper panel shows the distribution of all clumps (red) and the SEDIGISM GMC catalogue (blue).  In both cases, these curves show a steeply rising gradient at velocity offsets less than $\pm$10\,\kms, revealing that both clumps and clouds are tightly correlated with the spiral arms. We note that the clumps are more tightly correlated  with the arms than the clouds (the KS-test gives a $p$-value $\ll 0.003$), suggesting that clouds associated with dense clumps are more likely to be associated with the spiral arms.

The spiral-arm loci are derived from pulsar dispersion measurements and are independent of gas: the strong correlation between the molecular gas and the location of the arms is then quite significant. If we assume that all sources located within $\pm$10\,\kms\ of a spiral-arm locus are associated with an arm, we find that 65\,per\,cent of the clumps and 61\,per\,cent of clouds are so associated. We looked for differences in the velocity distribution between clumps associated with high-mass star formation tracers and the rest of the matched clouds with high-reliability matches, but found no significant difference from the complete clump and cloud populations.

The lower panel of Fig.\,\ref{fig:arm_offset_vlsr} illustrates the velocity offset distribution for all clouds and those associated with dense clumps (grey and red histograms, respectively). Although we find the clumps are slightly more tightly correlated with the spiral arms, we do not find a significant difference in the distributions of these clouds associated with clumps with those that are unassociated with clumps ($p$-value = 0.034), and so any difference between these two subsamples is likely to be quiet subtle. The strong peak of the clouds at zero offset reveals a tight correlation between the $\ell- v$ distribution of clouds and the loci of the \citet{taylor1993} model of the spiral arms. We have performed the same analysis using the rotation curve of \citet{brand1993} to determine the loci of the spiral arms, but found no significant difference between the distribution of the clouds with respect to the spiral arms. This distribution extends out to a velocity of $\sim$30\,\kms, and a Gaussian fit to the full cloud catalogue shown in the lower panel of Fig.\,\ref{fig:arm_offset_vlsr} gives $\sigma$ of $\sim9$\,\kms.  This is similar to the velocity dispersion of clouds from the velocity expected for circular rotation reported by  \citet{brand1993} (12.8\,\kms) and streaming due to the motion of clouds through the spiral arms ($\sim$7-10\,\kms; \citealt{burton1971,stark1989,reid2009}). 

\subsection{Dense gas fraction}

In this and the following subsection we will {\bf restrict} our analysis to clouds associated with dense gas that are located between $\ell = 300\degr$ and 350\degr\ in order to exclude the Galactic centre region, where the spiral-arm loci are extremely poorly constrained. We also exclude clouds and their associated clumps that are truncated at the edges of the survey region ($|b| \approx 0.5\degr$) as source properties there are less reliable. This reduces the cloud sample from 1695 associated with an ATLASGAL clump to 936.

\citet{urquhart2018} determined the clump masses and luminosities for all ATLASGAL clumps for which aperture photometry could be reliably extracted. We use these parameters together with the associations between clumps and their host GMCs provided by Paper\,III to calculate the dense-gas fraction (DGF$_{\rm gmc}=\sum M_{\rm clump}/M_{\rm gmc}$) and the instantaneous star-forming efficiency (i.e., SFE$_{\rm gmc}= \sum L_{\rm clump}/M_{\rm gmc}$).  These two parameters, as defined, are independent of heliocentric distance. The distances of clumps and the GMCs have been determined using different rotation curves and velocities so distances of matched objects can be different ($\pm0.5$\,kpc). We can eliminate this by first dividing the catalogue masses and luminosities by the catalogue distance squared to obtain cloud and clump properties per kpc$^2$ before taking the ratios of clump and cloud values (i.e. all normalised to 1\,kpc). 

The clump masses and cloud masses have been calculated using different dust opacity values. The ATLASGAL survey uses an opacity of 1.85\,cm$^{2}$\,g$^{-1}$ at $\nu_0 = 350$\,GHz (\citealt{schuller2009_full} and references therein) while the SEDIGISM survey uses a CO-H$_2$ conversion (CO X factor $\sim1.08\pm 0.19\times 10^{21}$\,cm$^{-2}$ (K\,km\,s$^{-1}$)$^{-1}$; see Paper\,I for details) that is derived from the dust opacity used by the Hi-GAL column density maps (\citealt{elia2013,schisano2020}; $\kappa_0 = 0.1$\,cm$^{2}$\,g$^{-1}$ at $\nu_0 = 1200$\,GHz; \citealt{hildebrand1983}). Following the reference of \citet{elia2013} the opacity is assumed to scale as:






\begin{equation}
\kappa_{\nu} = \kappa_0 \left( \frac{\nu}{\nu_0} \right)^\beta
\end{equation}

If we use the values used by \citet{elia2013} stated above and set $\nu = 350$\,GHz and set $\beta = 1.75$ we obtain a value of $\kappa_{\rm 350\,\mu m} = 8.5 \times 10^{-3}$\,cm$^{2}$\,g$^{-1}$ for the dust opacity at the ATLASGAL frequency. The difference in the dust results in a discrepancy of a factor of 1.61 in the masses and to compensate for this we have multiplied the ATLASGAL clump masses by this factor. We also need to apply this factor to the 5$\sigma$ column densities threshold for the ATLASGAL survey $\sim7.5\times 10^{21}$\,cm$^{-2}$ at 20\,K (\citealt{schuller2009_full}), which is now $\sim1.5\times 10^{22}$\,cm$^{-2}$. The column density sensitivity of SEDIGISM is $\sim0.95\times 10^{21}$\,cm$^{-2}$, assuming a 5$\sigma$ noise of 3.5\,K, channel width of 0.25\,\kms\ and the mean CO X factor. The material traced by ATLASGAL has column densities approximately 15 times larger than the GMCs identified by SEDIGISM.

\begin{figure}
\centering 
\includegraphics[width=0.49\textwidth]{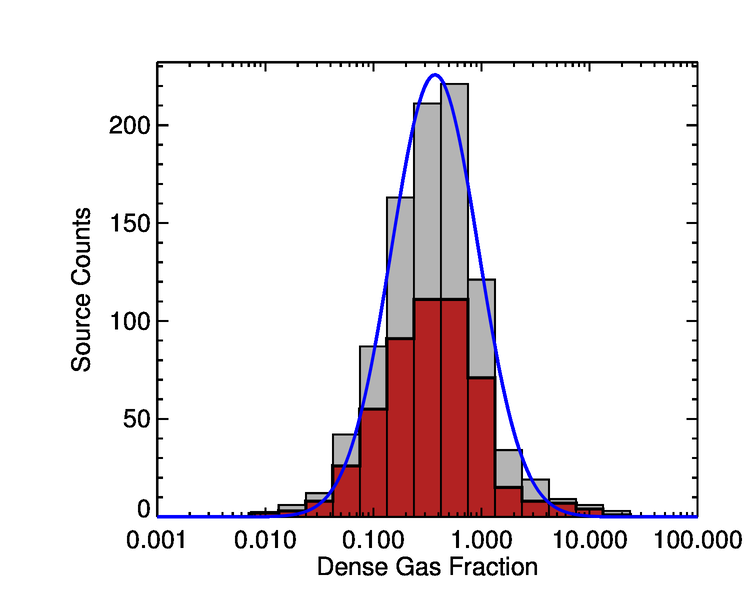}
\includegraphics[width=0.49\textwidth]{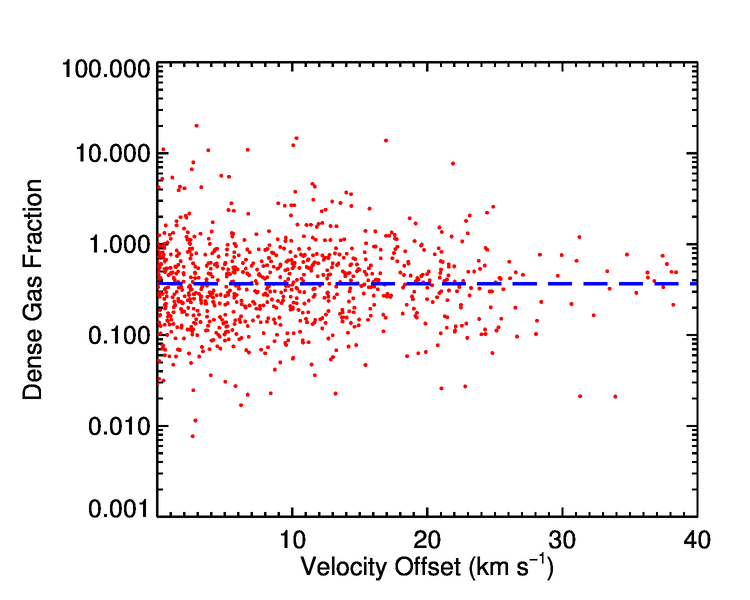}

\caption{Distribution of the GMC dense-gas fraction (DGF) determined by summing the masses of all embedded clumps and dividing by the GMC mass. The upper panel shows the distribution of the whole GMC sample (grey histogram) while the red histogram shows the distribution of clouds located within 10\,\kms\ of a spiral arm. The lower panel shows the distribution of DGF of each cloud as a function of velocity offset from their nearest spiral arms. The dashed blue line shows the results of a linear least-squares fit to the $\log_{10}$[DGF] and the velocity offset; this has a slope of $(-1.8 \pm 0.97) \times 10^{-3}$, which is essentially flat. The bin size used in the upper panel is 0.25\,dex. The blue curve is the result of a log-normal fit to the distribution of the whole GMC sample.} \label{fig:dgf_hist}
\end{figure}

In Fig.\,\ref{fig:dgf_hist} we show the distribution of the DGF with respect to the velocity offset between the clouds and their nearest spiral arms. We note that there is a large scatter in the values of the DGF (dex = 0.4) and this results in approximately 20\,per\,cent of the clouds having a DGF value above 1. Values of the DGF above unity are somewhat unexpected and might suggest the presence of an observational bias such as line-of-sight projection effects that might result in the dust continuum emission being systematically pushed to higher values. However, the spatial filtering of the large-scale dust emission that occurs as part of the data reduction effectively removes any contribution from large-scale diffuse emission and the overall number of dense clumps means that the chance of multiple clumps lying along the same line-of-sight is negligible. An alternative explanation for these high values for the DGF is the large uncertainties in the parameters that go into the calculation of the clump and cloud masses (i.e., CO X-factor and dust opacity). These values are only reliable to a factor of a few and taking the ratio of these masses can magnify the uncertainties and result in the large spread seen in Fig.\,\ref{fig:dgf_hist}. To illustrate this, let us consider the combined uncertainty in the ratio if the masses of the clumps are reliable to a factor of 2; adding the fractional uncertainties together in quadrature, we find that the total uncertainty is a factor of 2.5 or dex of 0.45, which is similar to what is observed here and in the studies by \citet{rigby2019} and \citet{eden2012,eden2013}. 

The uncertainties in the CO X-factor and dust opacity, which are thought to be constrained to within a factor of a few, will affect the masses of each population in a similar way, and so when combined they may lead to systematic shifts in the DGF up or down by a factor of a few, while variations in these values from cloud to cloud will also contribute to the spread in the distribution. Studies of the DGF have reported a range of mean values from 0.05 from a combination of ATLASGAL and PLANCK data (\citealt{csengeri2016_planck}) to 0.7 using different density thresholds on GRS data (\citealt{dib2012}) and so the choice of tracers and thresholds also plays a role. The absolute value of the DGF is therefore unlikely to be particularly reliable, however, trends in the distribution will be more robust.

The upper panel of Fig.\,\ref{fig:dgf_hist} shows the distribution of the DGF for all clouds associated with an ATLASGAL source (grey histogram) and the same sample of clouds that are within 10\,\kms\ of a spiral arm (red histogram). It is clear from this plot that there is no significant difference between these two populations (a KS-test gives a $p$-value of 0.06, which is a bit less than 2-sigma). This is consistent with the results reported by \citet{dib2012}, \citet{eden2012,eden2013}; these will be discussed in detail in Sect.\,\ref{sect:discussion}.

The mean value for the DGF is 0.35, which is considered to be an upper limit. This is in excellent agreement with the mean value determined from the ratio of ATLASGAL masses to CHIMPS clouds reported by \citet{rigby2019}. It is useful, at this point, to remember that only 11-17\,per\,cent of GMCs identified in SEDIGISM (Paper\,III) are associated with an ATLASGAL clump. We can obtain an estimate for the global DGF by considering the total mass of clumps and GMCs located between 300\degr\ and 350\degr\ in longitude, yielding a value of 0.16.  This is also likely to be an upper limit as the extraction algorithms used to identify coherent structures in the data cubes are only able to allocate approximately two-thirds of the $^{13}$CO emission to clouds (e.g. \citealt{rathborne2009,barnes2016}). The global DGF is therefore significantly lower than the upper limit derived here. It is also interesting to note that the overall shape of the distribution is log-normal distribution (the blue curve overplotted on Fig.\,\ref{fig:dgf_hist} shows the result of a log-normal fit to the full sample: a KS-test gives a $p$-value of 0.59 indicating there is no significant difference between distribution of the data and a log-normal distribution). 
This could indicate that variations in the DGF  from cloud to cloud are the result of a collection of essentially random processes (\citealt{eden2015}), the effects of which are multiplicative and in this case the extreme sources would not be abnormal. The statistical properties of the DGF are given in Table\,\ref{tbl:statistical_properties}.

The lower panel of  Fig.\,\ref{fig:dgf_hist} shows the DGF distribution as a function of offset from the nearest spiral arm: this plot reveals there is no enhancement of dense gas in clouds with respect to their proximity to the spiral arms. A linear least-squares fit to the data returns a slope that is very close to zero ($-1.8 \pm 0.97 \times 10^{-3}$), confirming there is no significant correlation between these two parameters (see dashed blue line on the lower panel of Fig.\,\ref{fig:dgf_hist}).

The lack of any enhancement of the DGF with proximity to the spiral arms is also consistent with the results of  \citet{csengeri2016_planck}, who calculated the fraction of emission from dense gas by comparing the emission detected in the ATLASGAL survey to the total dust emission detected by the Planck space mission ($\sim$5\,per\,cent). They also refer to this as a dense gas fraction, but it is averaged over the whole line of sight and includes a contribution from diffuse material not associated with the star formation process.  This therefore represents a lower-limit, and so is similar to the global DGF referred to above.

\setlength{\tabcolsep}{3pt}
\begin{table}

\begin{center}\caption{Summary of physical properties of the whole population of clumps and the four evolutionary subsamples identified. In Col.\,(2) we give the number of clouds in each subsample, in Cols.\,(3-5) we give the mean values, the error in the mean and the standard deviation, in Cols.\,(6-8) we give the median and minimum and maximum values of the samples.}
\label{tbl:statistical_properties}
\begin{minipage}{\linewidth}
\small
\begin{tabular}{lc......}
\hline \hline
  \multicolumn{1}{l}{Parameter}&  \multicolumn{1}{c}{\#}&	\multicolumn{1}{c}{$\bar{x}$}  &	\multicolumn{1}{c}{$\frac{\sigma}{\sqrt(N)}$} &\multicolumn{1}{c}{$\sigma$} &	\multicolumn{1}{c}{$x_{\rm{med}}$} & \multicolumn{1}{c}{$x_{\rm{min}}$}& \multicolumn{1}{c}{$x_{\rm{max}}$}\\
\hline
\multicolumn{8}{c}{All}\\
\hline
Log[DGF] &936&-0.45&0.01 & 0.45 & -0.45 & -2.11 & 1.30\\
Log[SFE GMC]& 936&-0.27&0.02 & 0.75 & -0.21 & -2.58 & 1.94\\
Log[SFE CSC]&         2829&0.27&0.01 & 0.80 & 0.27 & -2.09 & 2.72\\
\hline
\multicolumn{8}{c}{\vlsr < 10\,\kms}\\
\hline
Log[DGF]&          513&-0.47&0.02 & 0.47 & -0.48 & -2.11 & 1.30\\

Log[SFE GMC]&513&-0.32&0.03 & 0.79 & -0.26 & -2.58 & 1.94\\

Log[SFE CSC ]&         1685&0.30&0.02 & 0.82 & 0.30 & -2.09 & 2.72\\
\hline\\
\end{tabular}\\

\end{minipage}

\end{center}
\end{table}

\setlength{\tabcolsep}{6pt}

\subsection{Star formation efficiency}

\begin{figure}
\centering 

\includegraphics[width=0.49\textwidth]{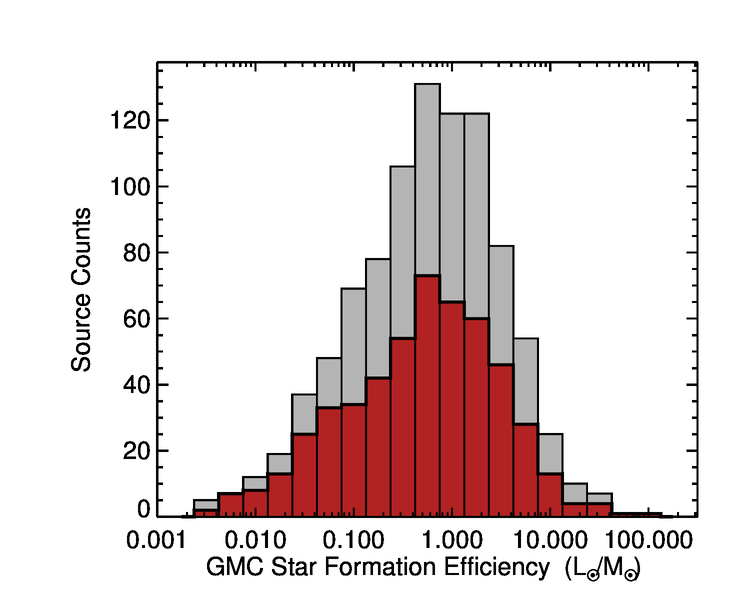}
\includegraphics[width=0.49\textwidth]{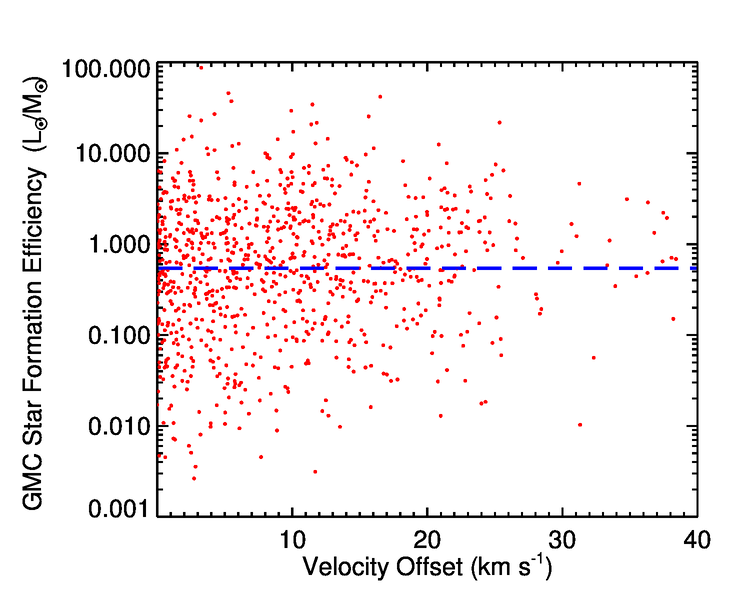}

\caption{As Figure\,\ref{fig:dgf_hist} but for the star formation efficiency.  The dashed blue line shows the results of a linear least-squares fit to the $\log_{10}$[SFE] and the velocity offset; this has a slope of $6.9\times 10^{-5}\pm 1.7\times 10^{-3}$, which is essentially flat. The bin size used in the upper panel is 0.25\,dex. }\label{fig:sfe_hist}
\end{figure}

Figure\,\ref{fig:sfe_hist} shows the results of our SFE calculation (i.e., ${\rm SFE_{gmc} }= \sum L_{\rm clump}/M_{\rm gmc}$) for the matched clumps and clouds located between $\ell = 300\degr$ and 350\degr.  In considering that $\sum L_{\rm clump}/M_{\rm gmc}$ is proportional to the SFE, we are implicitly assuming that the initial mass function (IMF) is universal and completely sampled. The universality of the IMF, or lack of it, are still highly debated, with some groups finding evidence of IMF variations in the MW (e.g., \citealt{dib2017}) and others arguing for a universal IMF (\citealt{bastian2010}). The results of \citet{dib2017} suggest that the slope at the high mass in the MW may have a standard deviation of $\approx$ 0.6 (i.e., IMFs with a slope as shallow as 0.7, and as steep as 2 around the Salpeter value of 1.35 can be found in the Milky Way), which is corroborated by other studies on smaller samples of clusters in the MW and M31 (e.g. \citealt{dib2014} and  \citealt{weisz2015}). How much variations in the IMF, stochastic or intrinsic, are affecting our own observations remains unclear, and is presently difficult to fold in our interpretation of the ($L/M$) ratio. Therefore, we make the assumption that the IMF is universal, as a working hypothesis.

The upper panel of Fig.\,\ref{fig:sfe_hist} shows the SFE distribution of all clouds (grey) and the subsample located within 10\,\kms\ of a spiral arm. The distributions of these two populations are very similar to each other, and a KS-test confirms that they are not significantly different ($p$-value of 0.79). The lower panel of Fig.\,\ref{fig:sfe_hist} shows the SFE as a function of velocity offset of the clouds from their nearest spiral arm. This plot shows no evidence for any significant change in the ${\rm SFE_{gmc} }$ as the velocity difference increases between the GMC and the spiral arms. A linear least-squares fit to the velocity offset and $\log_{10}$[SFE] has zero gradient ($6.9\times 10^{-5}\pm 1.7\times 10^{-3}$; see blue dashed line on the lower panel of Fig.\,\ref{fig:sfe_hist}), indicating that there is no correlation between these two parameters. From this we can conclude that there is no dependence of the star formation efficiency on a cloud's proximity to a spiral arm.

\begin{figure}
\centering 

\includegraphics[width=0.49\textwidth]{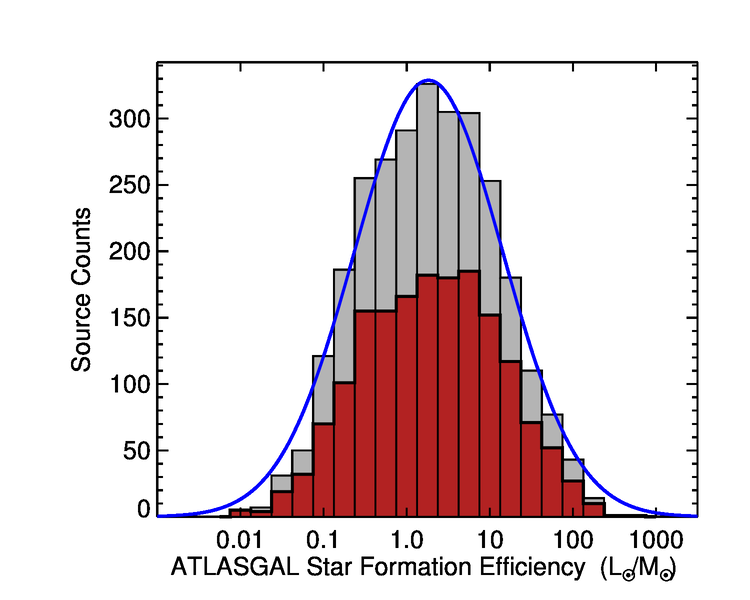}
\includegraphics[width=0.49\textwidth]{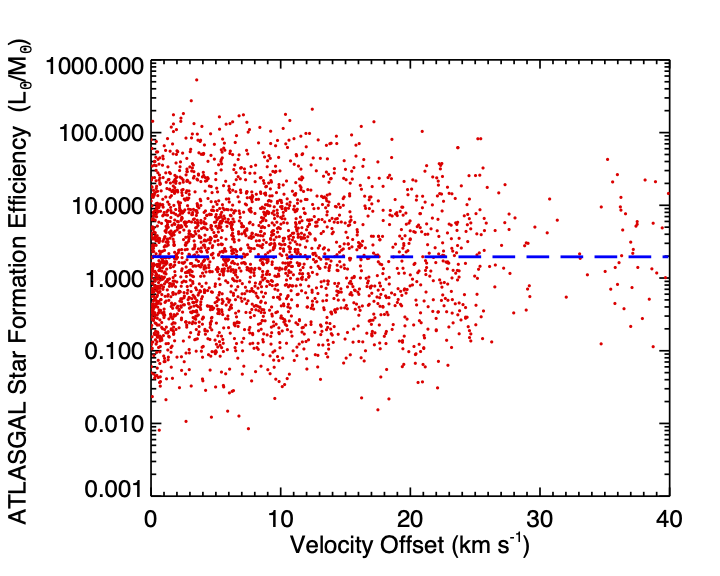}

\caption{As Figure\,\ref{fig:dgf_hist} but for the ATLASGAL clump star formation efficiency. The dashed blue line shows the results of a linear least-square fit to the Log$_{10}$[SFE$_{\rm clump}$] and the velocity offset; this has a slope of $(9.1\pm 1.5) \times 10^{-4}$,
which is essentially flat. The bin size used in the upper panel is 0.25\,dex. }\label{fig:csc_sfe}
\end{figure}

For completeness, we also provide plots of the star formation efficiency within individual ATLASGAL clumps (as calculated in \citealt{urquhart2018}) in Fig.\,\ref{fig:csc_sfe}. The distribution of the clump SFE is well modeled by a log-normal distribution (the log-normal fit to the data is shown by the blue curve overplotted on Fig.\,\ref{fig:csc_sfe}; a KS-test gives a $p$-value of 0.97 indicating there is no significant difference between the them) and its mean value is approximately an order of magnitude larger than the GMC SFE but the fraction associated with the spiral arms is approximately the same and has a similar distribution with respect to the spiral arms.

The statistical properties of the SFE for both the clumps and GMCs are given in Table\,\ref{tbl:statistical_properties}.

\subsection{DGF and SFE as a function of Galactocentric distance}

We have used the radial velocities obtained from the CO analysis in combination with the rotation curve of \citet{eilers2019} and the \citet{reid2019} values for Solar position and velocity (8.15\,kpc and 236\,\kms, respectively) to calculate the distances to the ATLASGAL clumps. Distances determined for clouds located within the Solar circle in this way are not unique, suffering from the kinematic distance ambiguity (KDA). The two positions that give rise to the KDA are positioned equidistant about a tangent location, and additional data are required resolve the ambiguity.  The DGF and SFE, however, are distance-independent quantities, and as we are primarily interested in the distribution with respect to the Galactic Centre, we do not need to resolve the KDAs, as both positions are equidistant from the Centre. We can therefore investigate the distribution of a range of distance-independent parameters with respect to their position in the Galactic disk. The Galactocentric distance for each cloud is given in Col.\,4 of Table\,\ref{tbl:cloud_derived_parameters}.

\begin{figure}
\centering 
\includegraphics[width=0.49\textwidth]{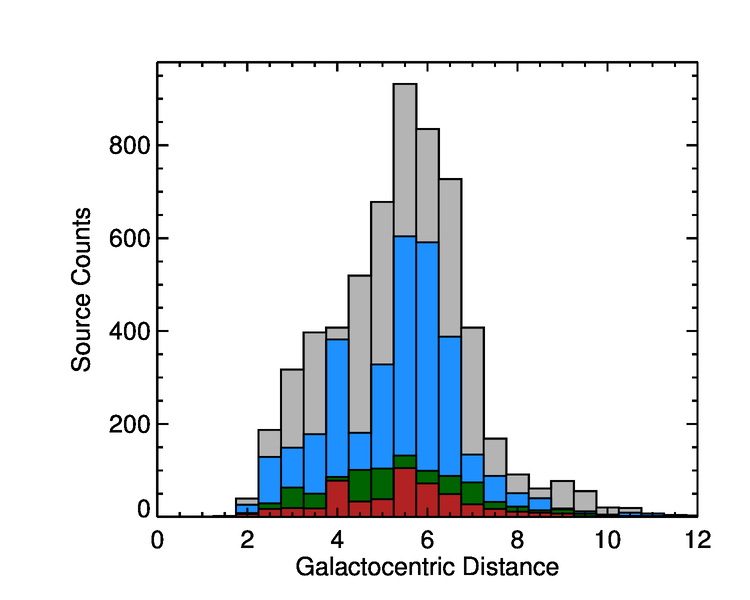}

\caption{Distribution of GMCs as a function of Galactocentric distance. The distribution of the whole sample is shown in grey, those located within 10\,\kms\ of a spiral arm, are shown in blue. The clouds associated with dense gas are shown in green and the population of these clouds that are within the 10\,\kms\ of a spiral arm are shown in red. The bin size used is 0.5\,kpc.}\label{fig:clumps_rgc}
\end{figure}

We show the distribution of all SEDIGISM GMCs (grey) and those located within 10\,\kms\ of a spiral arm (blue) in Figure\,\ref{fig:clumps_rgc}. While the overall distribution is relatively featureless, the latter is more structured, showing two strong peaks at Galactocentric distances of $\sim$4\,kpc and $\sim$5.75\,kpc.  The closer can be directly traced back via the $\ell v$-diagram to the near side of the Norma arm, and the further to the near side of the Scutum and far side of the Norma arms (but predominately the Scutum arm), respectively.
Figure\,\ref{fig:clumps_rgc} shows the distribution of all clouds associated with dense clumps (green) and those associated with both dense gas and a spiral arm (red). The clouds associated with both dense gas and spiral arms also show peaks at $\sim$4\,kpc and $\sim$5.75\,kpc and, importantly, these peaks incorporate nearly all clouds associated with dense gas in these distance bins (i.e. nearly all of the clouds in these bins can be attributed to the arms). 

\begin{figure}
\centering 

\includegraphics[width=0.49\textwidth]{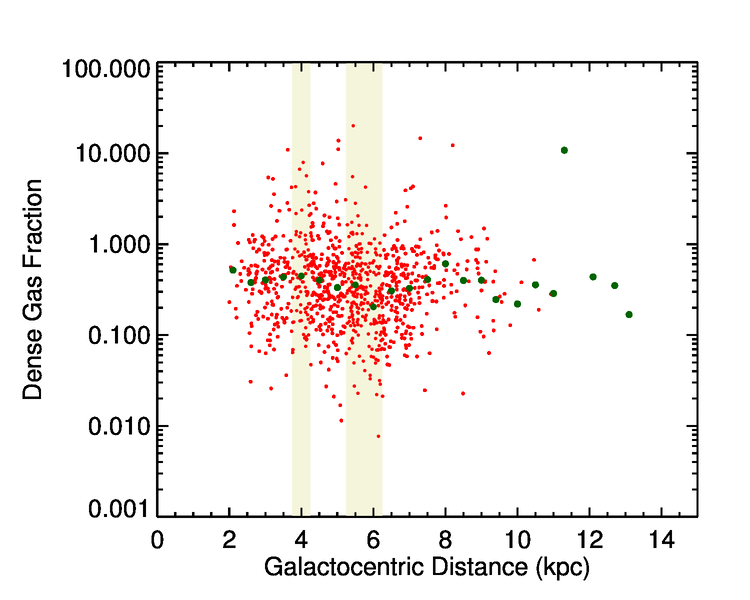}
\includegraphics[width=0.49\textwidth]{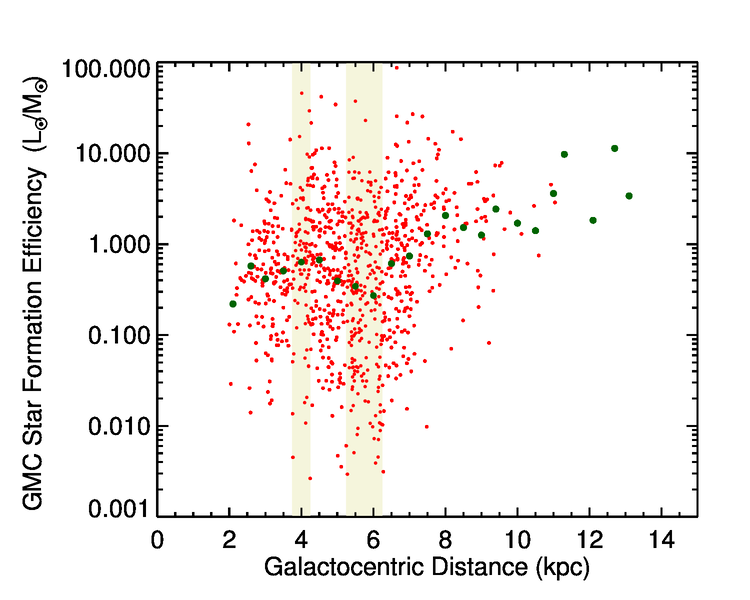}

\caption{Distribution of the GMC dense gas fraction and star formation efficiency as a function of distance from the Galactic centre. The vertical shaded areas on both plots correspond to regions in Galactocentric radius where the cloud population is dominated by the near sides of the Norma and Scutum arms (4\,kpc and 5.75\,kpc, respectively). The green circles indicate the average value calculated for each 0.5\,kpc.  }\label{fig:sfe_dgf_rgc}
\end{figure}

We show the DGF and SFE as a function of Galactocentric distance in the upper and lower panels of  Fig.\,\ref{fig:sfe_dgf_rgc}, respectively. The shaded vertical regions indicate the peaks seen in Fig.\,\ref{fig:clumps_rgc} that are associated with the Norma and Scutum arms, and this is where we would expect to see increases if these spiral arms were playing a role in enhancing these parameters. The distribution of the DGF is relatively flat across the disk and there is no evidence of any significant enhancements that might be attributable to spiral arms.

The distribution of the SFE measurements is a little more complicated, as there appears to be a positive correlation with increasing distance from the Galactic centre.  There is significantly more scatter in these measurements and there may be a bias towards clouds associated with more luminous star formation with increasing distance. If we ignore clouds outside of the solar circle (i.e., $R_{\rm gc} > 8.15$\,kpc) then the distribution in the inner disk looks relatively flat and again there is no evidence for significant localised increases in the SFE towards either the Norma or Scutum spiral arms.

As previously mentioned, \citet{dib2012} calculated the SFE and DGF for the GRS region ($\ell = 18\degr-55.7\degr$; \citealt{jackson2006}) and reported similarly flat distributions. Since the SEDIGISM survey was designed to complement the longitude coverage of the GRS and taken together we can conclude that there are no significant variations in the DGF or SFE across the whole of the inner part of the Galactic disc.

\section{Discussion}
\label{sect:discussion}
\subsection{Comparison with previous results}

The analysis presented in the previous section has failed to find any enhancement of the SFE or DGF with respect to a cloud's proximity to its nearest spiral arm. We do see significant amounts of cloud-to-cloud variation in these parameters, indicating that the clouds themselves have a wide range of physical and star-formation properties and that local initial conditions (i.e., local to the cloud formation site and scale) can have a significant impact on how efficiently dense clumps can be formed in individual GMCs.  While our analysis has focused on the properties of GMCs located in the 4$^{\rm th}$ Galactic quadrant, there has been a significant number of complementary studies that have focused on clouds located in the 1$^{\rm st}$ quadrant that we can take in context with our results to obtain a more global view.

Our failure to find any enhancement in the SFE in the vicinity of the spiral arms is supported by  studies by \citet{moore2012}  and \citet{dib2012}, both of which conducted similar analysis by matching up massive young stellar objects (MYSOs) and compact \hii\ regions identified by the Red MSX Source (RMS) survey (\citealt{lumsden2013,urquhart2014_rms}) with clouds identified by the Galactic Ring Survey (GRS;  for descriptions of the survey, and the cloud catalogue and cloud parameters see \citealt{jackson2006}, \citealt{rathborne2009}, \citealt{roman2009,roman-duval2010}, respectively).

\citet{moore2012} identified two peaks in the $L_{\rm RMS}$/$M_{\rm GRS}$ ratio at Galactocentric distances of 6 and 8\,kpc,  which correspond to the Sagittarius and Perseus spiral arms. These two peaks are dominated by the W49 and W51 star forming complexes (two of the  most extreme star forming complexes in the Galactic disk; \citealt{urquhart2018}) and when the contribution from these two regions was removed, the $L_{\rm RMS}$/$M_{\rm GRS}$ was found to be similar for both the arm and interarm regions. \citet{dib2012} also looked at the SFE as a function of distance from the spiral arms and found no significant correlation.

\citet{eden2012,eden2013} conducted a similar analysis using the molecular clouds extracted and parameterised from the GRS and the dense clumps identified by the Bolocam Galactic Plane Survey (BGPS; survey and catalogue description can be found in  \citealt{rosolowsky2010_bgps_ii, aguirre2011,ginsburg2013_bgps_ix}, while distances and physical properties can be found in \citealt{dunham2011_bgps_vii,shirley2013_bgps_x}). They refer to their parameter as the clump formation efficiency (CFE), but this is calculated in essentially the same way and so is consistent with our DGF. Eden et al. report a CFE of $14.9\pm4.8$ and $16.3\pm7.5$\,per\,cent for the spiral arm and interarm regions respectively, and conclude that there is no significant difference between these regions. This is similar to the DGF we have determined from our data ($\sim16$\,per\,cent when taking into account the whole GMC population) and fully consistent with our findings.

The work by \citet{eden2012,eden2013} and \citet{moore2012} used different tracers to map the molecular clouds and dense gas and to measure the luminosities, and focused on structures located in the 1$^{\rm st}$ Galactic quadrant.  As a result, their analyses provide an excellent complement to the analysis presented in this paper. Our work extends their analysis into the 4$^{\textrm{th}}$ Galactic quadrant and, when combined, both studies  cover a substantial fraction of the inner Galactic disk where the vast majority of $^{12}$CO (1-0)-traced molecular gas 
in the Galaxy resides \citep[i.e. $\sim$85\% within the Solar circle;][]{miville2017}. As such, we may conclude that there is no evidence that the spiral arms play a significant role in enhancing either the efficiency of converting molecular gas into dense clumps where star formation is known to be taking place, or the efficiency with which the GMCs are forming stars. 

These results are further supported by a recent study by \citet{ragan2016,ragan2018} which looked at the star formation fraction (SFF), defined as the ratio of the number of Hi-GAL clumps associated with a 70-\mum\ counterpart divided by the total number of HiGAL clumps in a fixed area across the whole inner Galactic disk ($-71\degr \le \ell \le 67\degr$). The 70\,\mum\ counterpart is used as an indicator of embedded star formation \citep{dunham2008,ragan2012,elia2017}.  That study found no significant enhancements in the prevalence of star formation as measured by the SFF across the disk, which is consistent with the lack of variation in the SFE$_{\rm gmc}$. These results are also consistent with the relatively flat SFE profiles seen on kpc scales between the centre and disks in the vast majority of nearby galaxies (e.g. \citealt{utomo2017}) and the simulations of spiral galaxies presented by \citet{kim2020} who found that, although 90\,per\,cent of the star formation is localised to spiral arms, the overall enhancement in the arms is less than a factor of two compared to the inner-arm regions.

\subsection{Role of the spiral arms}

Observations of nearby spiral galaxies clearly show that there are enhancements in the surface density of star formation activity in the  spiral arms \citep[e.g., M51,][]{hughes2013}. Our analysis of the $\ell v$-distribution of the GMCs and spiral-arm loci has revealed that the molecular gas in the Milky Way is tightly correlated with the spiral arms. \citet{rigby2019} has reported increases in velocity dispersion and the virial parameter in CHIMPS clumps associated with spiral arms. Additionally, peaks in the clump and cloud distributions as a function of Galactocentric distance can be attributed to specific spiral arms in both the 1$^{\rm st}$ and 4$^{\rm th}$ quadrants (this paper and \citealt{urquhart2018, moore2012, eden2013}). These locations have also been linked to peaks in the star-formation activity (\citealt{urquhart2014_rms}) and this is consistent with observations of nearby spiral galaxies, where the arms are seen to be both rich in dense molecular gas and in star formation (e.g. \citealt{helfer2003, hughes2013}). For all of these observations, however, neither this study nor similar studies reported in the literature have found any significant enhancements in the DGF or SFE attributable to the spiral structure. 

This leads us to conclude that the arms are principally collecting material together via orbit crowding but there is no evidence that they are playing a role in enhancing the star formation within molecular clouds. The increase in the star formation density found in the vicinity of spiral arms (e.g. \citealt{moore2012, urquhart2014_rms}) is likely to be the result of source crowding and not the result of any direct influence of the spiral arms themselves. This conclusion is supported by a recent study by \cite{Pettitt2020} that looked at star-forming regions in simulations of barred and armed Milky Way analogues, comparing them to the measurements from the Hi-GAL survey. They found only minor increases in star formation activity as a function of radius for the more Milky Way-like configurations (4-armed and barred disks). Their study looked at the Galaxy as a whole rather than just a single quadrant, so it is possible that radial signatures may appear more washed out in the disk averaging.

As a further component, in a recent study of the atomic gas in the northern Milky Way with the HI/OH/recombination line survey of the Milky Way (THOR; \citealt{beuther2016}), \citet{wang2020} found that the ratio of molecular to atomic gas changes by approximately a factor of six from the spiral arm to the interarm regions. Taken together, the picture that emerges from all of these studies is one in which the spiral arms do indeed play an important role in collecting gas, converting it from atomic to molecular gas,  and forming GMCs (\citealt{koda2016}), but  the subsequent star formation processes depend more on local effects, and mainly internal conditions, that may be the result of a combination of random processes.

\section{Summary and Conclusions}
\label{sect:summary}

We have used the first full data release from the SEDIGISM CO survey of the Galactic mid-plane between $300\degr \le \ell \le 18\degr$ (Paper\,II) to extract $^{13}$CO and C$^{18}$O ($J = 2-1$) line spectra towards dense clumps drawn from the ATLASGAL compact source catalogue (CSC; \citealt{contreras2013, urquhart2014_csc}) located in the same longitude range. These have been fitted with Gaussian profiles to obtain velocities, FWHM and peak intensities. The most appropriate velocity component and corresponding physical properties are assigned to each ATLASGAL clump, where ``appropriate'' is determined as one that is at a minimum a factor of two brighter than other components if multiple components are detected, or from a visual comparison between the morphologies of the CO emission and the dust emission. Useful spectra have been extracted towards 5148 clumps, and a reliable velocity has been assigned to 4998 clumps. This adds 1108 clumps for which a velocity was not previously available, and corrects velocities for a further 269 sources.  
We have used a catalogue of GMCs that have been matched to the ATLASGAL clumps (Paper\,III) to identify the parental molecular clouds for the dense clumps located in the 4$^{\rm th}$ quadrant and determine their star-forming properties. Only a small fraction of GMCs are associated with dense gas (11\,per\,cent) and these tend to be the more massive and larger GMCs. We have calculated the dense-gas fraction (DGF$_{\rm gmc}=\sum M_{\rm clump}/M_{\rm gmc}$) and the instantaneous star-formation efficiency (SFE$_{\rm gmc} = \sum L_{\rm clump}/M_{\rm gmc}$) for the host GMCs by summing up the mass and luminosities of the embedded clumps  (determined by \citealt{urquhart2018}) and dividing this by the GMC's mass (Paper\,III).  We use these distance-independent quantities to look for variations with proximity to spiral arms that might provide some insight into their role in the star formation process. We have been able to put limits on the range of global dense gas fraction at 5-16\,per\,cent.

Our analysis of the velocity differences between GMCs and their nearest spiral arm has revealed that the vast majority are located within 20\,\kms\ of a spiral arm. The velocity offset distribution is strongly peaked at zero\,\kms\ but decreases smoothly out to 20\,\kms, indicating that the spiral arms are not particularly well defined in velocity. This may indicate that the spiral structure of the Milky Way is more flocculent than Grand Design. We also looked at the variations in these quantities as a function of their proximity in velocity to the spiral-arm loci and at specific Galactocentric distances where we expect the population to be dominated by the Norma and Scutum spiral arms. 

Neither of these two methods has found evidence of a significant increase in either the DGF or SFE with respect to a cloud's proximity to a spiral arm. These results are consistent with the results of similar independent studies focusing on clouds located in the 1$^{\rm st}$ quadrant of the Galaxy and, combined, provide strong evidence that the spiral arms do not enhance either the formation of dense gas in molecular clouds or the star-formation efficiency. Although our analysis has not found any evidence for large-scale influence in the $\rm DGF_{ gmc}$ and $\rm SFE_{ gmc}$, we have noted the presence of significant cloud-to-cloud variations in these parameters.  We have attributed these small-scale variations to differences in the environmental conditions indicating that the star-formation conditions within individual clouds can vary a great deal. 

The spiral arms are important for collecting material and for converting \hi\ to H$_2$ but play little part in the subsequent formation of dense clumps or their collapse into stars. The increase in star formation normally found towards the spiral arms is, therefore, likely to be the result of source crowding within the arms and not due to any direct influence from the arms themselves.

\section*{Acknowledgments}

We would like to thanks the referee for their comments and suggestions that have help to improve the clarity of a number of important points. This document was produced using the Overleaf web application, which can be found at www.overleaf.com. T.C. has received financial support from the French State in the framework of the IdEx Université de Bordeaux Investments for the future Program. H.B. acknowledges support from the European Research Council under the Horizon 2020 Framework Program via the ERC Consolidator Grant CSF-648505. H.B. also acknowledges support from the Deutsche Forschungsgemeinschaft (DFG) via Sonderforschungsbereich (SFB) 881 “The Milky Way System” (sub-project B1). L.B acknowledges support from CONICYT project Basal AFB-170002. SL, AG and ES have been supported by INAF through the project Fondi mainstream `Heritage of the current revolution in star formation: the Star-forming filamentary Structures in our Galaxy'. The work has partly been carried out within the Collaborative Research Centre 956, sub-project A6, funded by the Deutsche Forschungsgemeinschaft (DFG) - project ID 184018867. This project has received funding from the European Union's Horizon 2020 research and innovation program under grant agreement No. 639459 (PROMISE) and (MW) the Marie Skłodowska-Curie grant agreement No 796461.

\section*{Data Availability}

The data underlying this article are available in the article and in its online supplementary material. The observational datasets this paper is based on are available for download from the ATLASGAL\footnote{https://atlasgal.mpifr-bonn.mpg.de/cgi-bin/ATLASGAL\_DATABASE.cgi} and SEDIGISM\footnote{https://sedigism.mpifr-bonn.mpg.de/cgi-bin-seg/SEDIGISM\_DATABASE.cgi} survey web servers. 

\bibliographystyle{mnras}
\bibliography{urquhart_2019,references}


\bigskip

\section*{Affiliations}

\noindent $^{1}$ Centre for Astrophysics and Planetary Science, University of Kent, Canterbury, CT2\,7NH, UK \\
{\color{black}$^{2}$ Wartburg College, Waverly, IA, 50677, USA}\\
{\color{black}$^{3}$ Astrophysics Research Institute, Liverpool John Moores University, Liverpool Science Park Ic2, 146 Brownlow Hill, Liverpool, L3\,5RF, UK}\\
$^{4}$ School of Physics and Astronomy, Cardiff University, Cardiff CF24 3AA, UK \\
$^{5}$ Department of Physics, Faculty of Science, Hokkaido University, Sapporo 060-0810, Japan\\
$^{6}$ Max-Planck-Institut f\"ur Radioastronomie, Auf dem H\"ugel 69, D-53121 Bonn, Germany \\
$^{7}$ Leibniz-Institut f\"ur Astrophysik Potsdam (AIP), An der Sternwarte 16, 14482 Potsdam, Germany \\
$^{8}$ Laboratoire d'astrophysique de Bordeaux, Univ. Bordeaux, CNRS, B18N, all\'ee Geoffroy Saint-Hilaire, 33615 Pessac, France \\
$^{9}$Laboratoire d’Astrophysique (AIM), CEA, CNRS, Université Paris-Saclay, Université Paris Diderot, Sorbonne Paris Cité, 91191 Gif-sur-Yvette, France\\
$^{10}$ Max-Planck-Institut f\"ur Astronomie, K\"onigstuhl 17, D-69117 Heidelberg, Germany \\
$^{11}$ West Virginia University, Department of Physics \& Astronomy, P. O. Box 6315, Morgantown, WV 26506, USA \\
$^{12}$ Adjunct Astronomer at the Green Bank Observatory, P.O. Box 2, Green Bank WV 24944\\
$^{13}$ Center for Gravitational Waves and Cosmology, West Virginia University, Chestnut Ridge Research Building, Morgantown, WV 26505\\
$^{14}$ Space Science Institute, 4765 Walnut Street, Suite B, Boulder CO 80301 USA\\
$^{15}$ INAF-Osservatorio Astrofisico di Arcetri, Largo Enrico Fermi 5, I-50125 Firenze, Italy	\\
$^{16}$ Departamento de Astronom\'ia, Universidad de Chile, Casilla 36-D, Santiago, Chile \\
$^{17}$ INAF - Istituto di Radioastronomia, Via Gobetti 101, 40129 Bologna, Italy \\
$^{18}$ Chalmers University of Technology, Department of Space, Earth and Environment, SE-412 93 Gothenburg, Sweden \\
$^{19}$ Haystack Observatory, Massachusetts Institute of Technology, 99 Millstone Road, Westford, MA 01886, USA \\
$^{20}$ Korea Astronomy and Space Science Institute
776 Daedeok-daero, Yuseong-gu, Daejeon 34055, Republic of Korea\\
$^{21}$ INAF - Osservatorio Astronomico di Cagliari, Via della Scienza 5, 09047 Selargius (CA), Italy \\
$^{22}$  European Southern Observatory, Alonso de Cordova 3107, Vitacura, Santiago, Chile\\
$^{23}$ I. Physikalisches Institut, Universit\"at zu K\"oln, Z\"ulpicher Str. 77, D-50937 K\"oln, Germany \\
$^{242}$ INAF – Istituto di Astrofisica e Planetologia Spaziali, Via Fosso del Cavaliere 100, 00133 Roma, Italy\\
$^{25}$ Department of Physics \& Astronomy, University of Exeter, Stocker Road, Exeter, EX4 4QL, United Kingdom \\

\end{document}